\documentclass[aps,amsmath,nofootinbib,amssymb,superscriptaddress]{revtex4}
\usepackage{amssymb,graphics,graphicx,color,epstopdf,subfigure,dcolumn,bm,slashed,amsmath}
\usepackage{fancyhdr}

\def\eps{\epsilon}

\begin{document}

\begin{picture}(15,65)(20,-150)
 \put(400,-100){KEK-TH-2025}
   \put(400,-110){CP3-17-62}
 \put(400,-120) {KIAS-Q18001}
  \put(400,-130){OU-HET-959}
\end{picture}

\title{Probing the CP properties of top Yukawa coupling at an $e^+e^-$ collider}

\author{Kaoru Hagiwara} \email{kaoru.hagiwara@kek.jp}
\affiliation{KEK Theory Center and Sokendai, Tsukuba, Ibaraki 305-0801, Japan}
\affiliation{Centre for Cosmology, Particle Physics and Phenomenology (CP3), Universit\'e Catholique de Louvain, B-1348 Louvain-la-Neuve, Belgium}
\author{Hiroshi Yokoya} \email{hyokoya@kias.re.kr}
\affiliation{Quantum Universe Center, KIAS, Seoul 02455, Korea}
\author{Ya-Juan Zheng} \email{yjzheng@het.phys.sci.osaka-u.ac.jp}
\affiliation{Department of Physics, Osaka University, Osaka 560-0043, Japan}

\date{\today}

\begin{abstract}

We study consequences of CP violation in the $ht\bar{t}$ Yukawa
 coupling through the process $e^+e^- \to$ $h(125)t\bar{t}$. %
 The helicity amplitudes are calculated in the $t\bar{t}$ rest frame,
 where the initial $e^+e^-$ current and the final Higgs boson have
 the same three-momentum.
 CP-violating asymmetries appear not only in the azimuthal angle
 between the $e^+e^-$ plane and the $t\bar{t}$ plane about the
 Higgs momentum direction, but also in the correlated decay
 angular distributions of $t$ and $\bar{t}$.
 Complete description of the production and decay angular
 distributions are obtained analytically, including both leptonic and hadronic decays of $t$ and $\bar{t}$. We study the
 ultimate sensitivity to the CP-violating $ht\bar{t}$ coupling at a few center-of-mass energies.
 Our analysis shows that the possibility of discovering CP-violating $ht\bar{t}$ coupling improves significantly by studying
 $t\bar{t}$ decay angular correlations, and more importantly,
 by increasing its energy upgrade target from $\sqrt{s} = 500$~GeV to 550~GeV. %
 
\end{abstract}

\maketitle

\section{Introduction}
All the current measurements of the Higgs boson $h(125)$ are consistent
with predictions of the Standard Model (SM) of particle
physics~\cite{Aad:2015gba,Khachatryan:2016vau}.  Detailed studies of all
its properties should be of at most importance in our search for the
physics of the symmetry breakdown, or the origin of the SM.

In this paper, we show how the top quark Yukawa coupling, the $ht\bar{t}$ coupling, can be studied in the process 
$e^+e^-\to ht\bar{t}$
at linear colliders such as ILC and CLIC. We develop techniques that allow us to perform the optimal measurements of the $ht\bar{t}$ coupling in a clean $e^+e^-$ collider environment. In order to quantify the impacts of our proposal to measure full production and decay angular correlations of both semi-leptonic and hadronic decays of top and anti-top quarks, we adopt the following simple effective Lagrangian for the $ht\bar{t}$ coupling
\begin{eqnarray}
{\cal L}=-g_{htt}h\bar{t}\left[\cos\xi_{htt}+i\sin\xi_{htt}\gamma_5\right]t\label{eq:lag}
\end{eqnarray}
with two real parameters, $g_{htt}$ and $\xi_{htt}$. The effective Lagrangian Eq.~(\ref{eq:lag}) reduces to the SM $ht\bar{t}$ coupling when 
\begin{eqnarray}
g_{htt}=g_{htt}^{\rm SM}=\frac{m_t}{v},~~\xi_{htt}=0
\end{eqnarray}
with the top quark mass $m_t$ and the vacuum expectation value (VEV) $v\simeq246$~GeV of the SM Higgs field. Nonzero values of $\sin\xi_{htt}$ term in Eq.~(\ref{eq:lag}) implies CP violation, while the magnitude of the $g_{htt}$ is measured with respect to its SM value
\begin{eqnarray}
\kappa_{htt}=g_{htt}/g_{htt}^{\rm SM}.
\end{eqnarray}
The sign of $\cos\xi_{htt}$ term, or that of $g_{htt}\cos\xi_{htt}$, is measured with respect to the sign of the $hZZ$ coupling, 
\begin{eqnarray}
g_{hZZ}^{}=\kappa_{hZZ}^{}~g_{hZZ}^{\rm SM}=\kappa_{hZZ}^{}~\frac{m_Z^2}{v},
\end{eqnarray}
because the amplitudes with the $ht\bar{t}$ couplings and those with the $hZZ$ coupling interfere in the process $e^+e^-\to ht\bar{t}$. Therefore, the amplitudes depend on the three parameters,
\begin{eqnarray}
\kappa_{htt},\kappa_{hZZ}>0, ~~~-\pi\leq\xi_{htt}<\pi.
\end{eqnarray}
It is worth noting that  by the time an $e^+e^-$ collider starts
studying the $ht\bar{t}$ production process, we should have constraints
on $(\kappa_{htt},\xi_{htt},\kappa_{hZZ})$ from the LHC: {$\kappa_{hZZ}$
can be measured from $\Gamma(h\to ZZ^\ast)$ and from the weak boson
fusion production cross section; $(\kappa_{htt},\xi_{htt})$ are measured
from $ht\bar{t}$ production process
{~\cite{Gunion:1996xu,Ellis:2013yxa,Demartin:2014fia, Khatibi:2014bsa,He:2014xla,Buckley:2015vsa}; and
$(\kappa_{htt},\xi_{htt},\kappa_{hZZ})$ can be constrained in the single top plus Higgs production process 
%where the amplitudes of the $ht\bar{t}$ and the $hZZ$ couplings interfere
~\cite{Biswas:2012bd,Ellis:2013yxa,Demartin:2015uha,Englert:2014pja}. Some CP violating asymmetries including top decay lepton distributions are studied in~\cite{Ellis:2013yxa}.
The couplings are constrained by the perturbative unitarity~\cite{Bhattacharyya:2012tj,Choudhury:2012tk}, and affect loop induced amplitudes for $\Gamma(h\to gg)$ and $\Gamma(h\to\gamma\gamma)$. }
The role of the $e^+e^-$ collider experiments should be their refinements and possible discovery of non-SM physics in the $ht\bar{t}$ coupling, such as CP violation.
{Differential cross sections and top polarizations are studied in~\cite{BarShalom:1995jb,BhupalDev:2007ftb}, including 
CP violating observables~\cite{BarShalom:1995jb}. Ref.~\cite{Hagiwara:2016rdv} studied asymmetries in decay lepton angular correlations in Higgs plus topponium production at $\sqrt{s}=500$ GeV.}}

At present, only the $hZZ$ coupling strength $\kappa_{hZZ}$ has been measured to be 
\begin{eqnarray}
\kappa_{hZZ}\gtrsim0.85
\end{eqnarray}
more or less free from detailed model
assumptions~\cite{Khachatryan:2016vau}. We expect the measurement to
improve significantly by the time of an $e^+e^-$ collider experiment,
and we set  
\begin{eqnarray}
\kappa_{hZZ}=1\label{eq:kzz1}
\end{eqnarray}
throughout our analysis. All the results are insensitive to this assumption, which does not change significantly by varying $\kappa_{hZZ}^{}$ in the $0.85<{\kappa_{hZZ}^{}}\leq1$ range, because the amplitudes with the $hZZ$ coupling are subleading in $e^+e^-\to ht\bar{t}$ process. We allow the two real parameters $(\kappa_{htt},\xi_{htt})$ to vary freely in our analysis, with the understanding that they should be constrained {significantly} by the LHC experiments when the $e^+e^-$ collider experiments are performed.

Before starting our studies, we find it instructive to examine a very
specific limit of CP violation in the Higgs couplings where the sole
origin of CP violation in the Higgs sector is in the Higgs potential or the Higgs self interactions, while all the other couplings including Yukawa couplings are CP conserving. It is like a milli-weak theory of CP violation for the neutral $K$ system~\cite{Wolfenstein:1964ks}, where CP violation is confined to the $K^0-\bar{K}^0$ mixing. The scenario can be realized in any multi-doublet models, where the observed Higgs boson, $h(125)$, is a linear combination 
\begin{eqnarray}
h(125)=O_{hH}H+O_{hH^\prime}H^\prime+O_{hA}A,
\end{eqnarray}
where $H$ and $H^\prime$ are CP-even while $A$ is a CP-odd neutral components of the Higgs boson. If all the Yukawa interactions of the current states, $H$, $H^\prime$ and $A$ are CP conserving, the sole origin of all the CP-violating couplings of $h(125)$ is the mixing matrix element $O_{hA}$. We can choose $H$ to be the fluctuation of the full SM Higgs VEV that gives the weak boson masses, and the orthogonal states $H^\prime$ and $A$ are fixed uniquely in two Higgs doublet models~(2HDM). Specific form of the $ht\bar{t}$ couplings is model dependent. For instance in type II 2HDM, the $hZZ$ and $ht\bar{t}$ couplings are 
\begin{eqnarray}
\kappa_{hZZ}&=&O_{hH}\nonumber\\
\kappa_{htt}\cos\xi_{htt}&=&O_{hH}+O_{hH^\prime}\frac{1}{\tan\beta}\nonumber\\
\kappa_{htt}\sin\xi_{htt}&=&O_{hA}\frac{1}{\tan\beta}
\end{eqnarray}
where $\tan\beta$ is the ratio of the two VEV's. It is clear that in this scenario, $\kappa_{hZZ}=1$ implies $O_{hH^\prime}=O_{hA}={0}$ from the orthogonality of the mixing matrix. Even in this scenario, significant CP violation is possible for the $ht\bar{t}$ coupling for small $\tan\beta$, if $\kappa_{hZZ}\sim$ 0.9.

The paper is organized as follows. In Section~\ref{sec:helicity}, we calculate the helicity amplitudes of the $e^+e^-\to ht\bar{t}$ process in the $t\bar{t}$ rest frame. In Section~\ref{sec:results}, we show the numerical results of production cross section, invariant mass  distribution of the $ht\bar{t}$ production process with next-to-leading-order (NLO) QCD corrections including Coulomb resummation for topponium formation. In Section~\ref{sec:chisq}, we introduce the density matrix formalism to express full kinematical distributions including $t$ and $\bar{t}$ decay angular correlations, including both semi-leptonic ($t\to b\bar{\ell}\nu$) and hadronic ($t\to b\bar{d}u$) decays. By accounting for uncertainties in quark jet-flavor identification, we study how the sensitivity to CP violation increases by measuring full distributions, at $\sqrt{s}=500$~GeV, 550~GeV and 1000~GeV. Summary and conclusions are given in Section~\ref{sec:sum}. In Appendix~\ref{sec:appA}, we show $t$ and $\bar{t}$ decay density matrices for both semi-leptonic and hadronic decays. In Appendix~\ref{sec:appB}, we review HELAS phase convention that affects our production density matrix elements. 

\section{Helicity amplitudes and density matrix}\label{sec:helicity}
\begin{figure}[htb]
\includegraphics[width=0.32\textwidth]{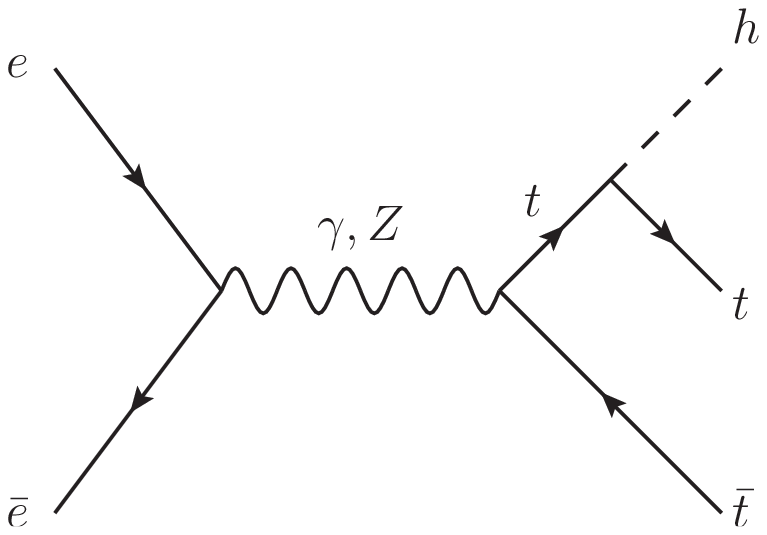}
\includegraphics[width=0.32\textwidth]{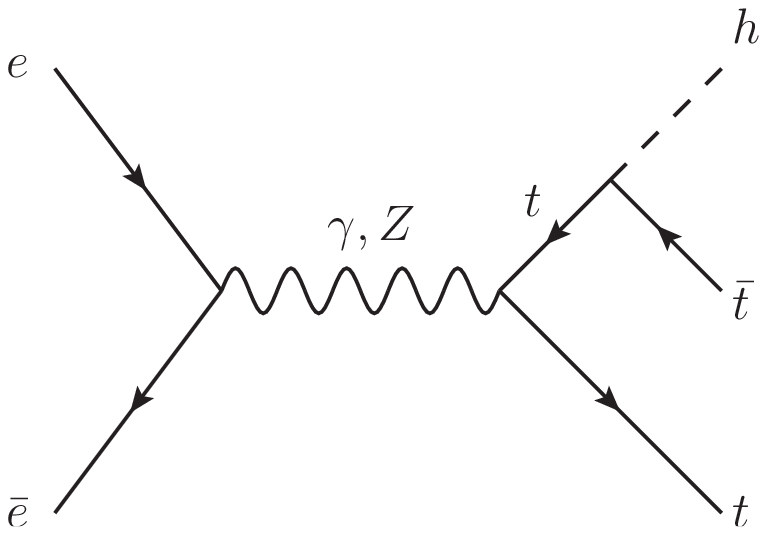}
\includegraphics[width=0.32\textwidth]{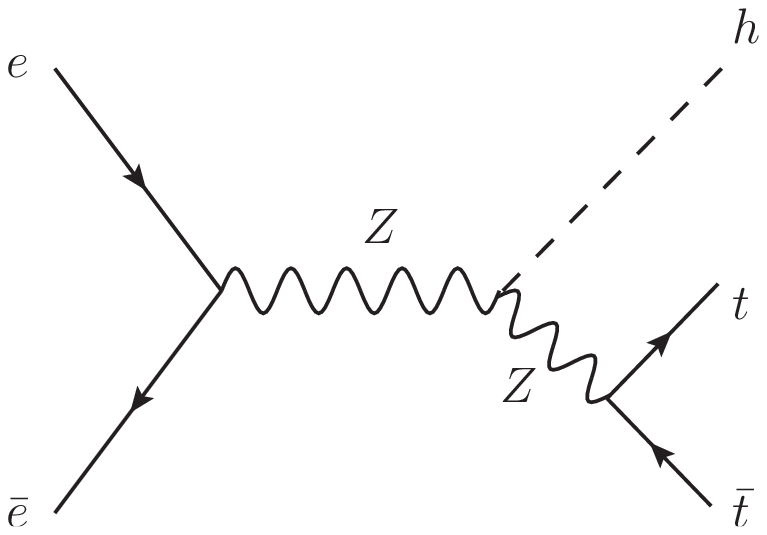}
\caption{
The Feynman diagrams of the $e\bar{e}\to ht\bar{t}$ process. 
 }\label{fig:feydiagram}
\end{figure}
As shown in Fig.~\ref{fig:feydiagram}, three Feynman diagrams contribute to the process $e\bar{e}\to ht\bar{t}$. The first two are with the $ht\bar{t}$ coupling in Eq.~(\ref{eq:lag}), and the third is with the $hZZ$ coupling for which we assume the SM {value} as in Eq.~(\ref{eq:kzz1}) in this report. 

\begin{figure}[t]
\includegraphics[width=0.45\textwidth]{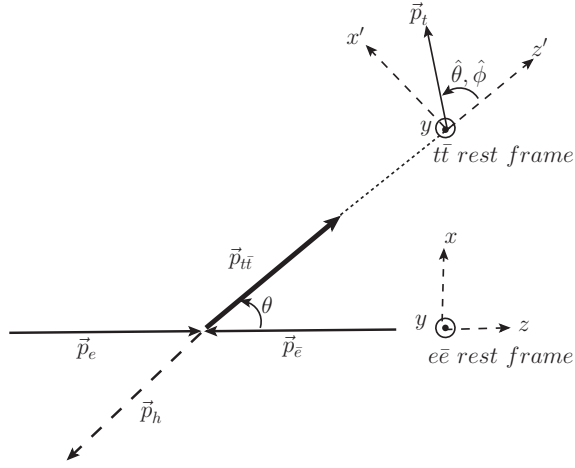}
\caption{Kinematics of the $e\bar{e}\to ht\bar{t}$ process in the $e\bar{e}$ rest frame and in the $t\bar{t}$ rest frame
 }\label{fig:kine}
\end{figure}

We calculate the helicity amplitudes of the process in the $t\bar{t}$ rest frame, where the initial $e^+e^-$ current and the produced Higgs boson have the same three-momentum. We denote the helicity amplitudes as 
\begin{eqnarray}
{\cal M}_{\alpha\sigma\bar{\sigma}}={\cal M}\left(e(p_e,\frac{\alpha}{2})+\bar{e}(p_{\bar{e}},-\frac{\alpha}{2})\to h(p_{h})+t(p_t,\frac{\sigma}{2})+\bar{t}(p_{\bar{t}},\frac{\bar{\sigma}}{2})\right),
\label{eq10:amp}
\end{eqnarray}
where $\alpha/2=\pm1/2$ gives the electron helicity, $\sigma/2$ and $\bar{\sigma}/2$ give the helicity of $t$ and $\bar{t}$ respectively in the $t\bar{t}$ rest frame. In order to fix our reference frame unambiguously, we start from the $e^+e^-$ collision (laboratory) frame, where the four momenta are parameterized as 
\begin{subequations}
\begin{align}
p_e^\mu&=\frac{\sqrt{s}}{2}(1,0,0,1),\\
p_{\bar{e}}^\mu&=\frac{\sqrt{s}}{2}(1,0,0,-1),\\
Q^\mu&=p_t^\mu {+} p_{\bar{t}}^\mu=\frac{\sqrt{s}}{2}\left(1+\frac{m_{t\bar{t}}^2-m_h^2}{s},\bar{\beta}\sin\theta,0,{\bar{\beta}}\cos\theta\right)=m_{t\bar{t}}\gamma(1,\beta\sin\theta,0,\beta\cos\theta),\\
p_h^\mu&=\frac{\sqrt{s}}{2}\left(1+\frac{m_h^2-m_{t\bar{t}}^2}{s},-\bar{\beta}\sin\theta,0,-\bar{\beta}\cos\theta\right)=(E_h,-p_h\sin\theta,0,-p_h\cos\theta).
\end{align}
\end{subequations}
Here $m_{t\bar{t}}=\sqrt{Q^2}$ is the invariant mass of the $t\bar{t}$ system,
\begin{eqnarray}
\bar{\beta}=\frac{1}{s}\lambda^{1/2}(\sqrt{s},m_{t\bar{t}},m_h),
\label{eq:betabar}
\end{eqnarray}
with $\lambda(a,b,c)=(a+b+c)(a+b-c)(a-b+c)(a-b-c)$, and
\begin{eqnarray}
\gamma=\frac{E_{t\bar{t}}}{m_{t\bar{t}}}=\frac{\sqrt{s}}{2m_{t\bar{t}}}\left(1+\frac{m_{t\bar{t}}^2-m_h^2}{s}\right),~~\gamma\beta=\frac{p_{t\bar{t}}}{m_{t\bar{t}}}=\frac{\sqrt{s}}{2m_{t\bar{t}}}\bar{\beta},
\label{eq:betagamma}
\end{eqnarray}
are the Lorentz boost factors between the $e\bar{e}$ and the $t\bar{t}$ rest frames. The $t$ and $\bar{t}$ momenta are parameterized in the $t\bar{t}$ rest frame which is obtained from the laboratory frame by a rotation of {$-\theta$} about the $y$-axis and then by a Lorentz boost along the $t\bar{t}$ momentum direction, which give
\begin{subequations}\label{eq:p}
\begin{align}
p_t^\mu&=\frac{m_{t\bar{t}}}{2}\left(1,\hat{\beta}\sin\hat{\theta}\cos\hat{\phi},\hat{\beta}\sin\hat{\theta}\sin\hat{\phi},\hat{\beta}\cos\hat{\theta}\right),\label{eq:pt}\\
p_{\bar{t}}^\mu&=\frac{m_{t\bar{t}}}{2}\left(1,-\hat{\beta}\sin\hat{\theta}\cos\hat{\phi},-\hat{\beta}\sin\hat{\theta}\sin\hat{\phi},-\hat{\beta}\cos\hat{\theta}\right),\label{eq:ptbar}\\\
p_h^\mu&=(\sqrt{s}\gamma-m_{t\bar{t}},0,0,-\sqrt{s}\gamma\beta),\label{eq:ph}\\
p_{e\bar{e}}^\mu&=(\sqrt{s}\gamma,0,0,-\sqrt{s}\gamma\beta),\label{eq:pee}
\end{align}
\end{subequations}
 with
\begin{eqnarray}
\hat{\beta}=(1-4m_t^2/m_{t\bar{t}}^2)^{\frac{1}{2}}.
\label{eq:betahat}
\end{eqnarray}
 The $z$-axis is chosen along the negative of the common momentum direction of the Higgs boson and the $e^+e^-$ system; see Fig.~\ref{fig:kine}. 
We find that the helicity amplitudes are most easily and systematically calculated in the frame obtained from the above by a rotation of $\hat{\phi}$ about $z$-axis, such that the top momentum has vanishing $y$-component:
\begin{eqnarray}
p_t^\mu&=&\frac{m_{t\bar{t}}}{2}(1,\hat{\beta}\sin{\hat{\theta}},0,\hat{\beta}\cos\hat{\theta}),\nonumber\\
p_{\bar{t}}^\mu&=&\frac{m_{t\bar{t}}}{2}(1,-\hat{\beta}\sin\hat{\theta},0,-\hat{\beta}\cos\hat{\theta}),
\end{eqnarray}
 the $h$ and $e^+e^-$ four momenta are unchanged from Eq.~(\ref{eq:ph}) and (\ref{eq:pee}), whereas the initial $e$ and $\bar{e}$ momenta are now,
\begin{eqnarray}
p_e^{\mu}&=&\frac{\sqrt{s}}{2}(\gamma(1-\beta\cos\theta),-\sin\theta\cos\hat{\phi},\sin\theta\sin\hat{\phi},\gamma(\cos\theta-\beta)),\nonumber\\
p_{\bar{e}}^\mu&=&\frac{\sqrt{s}}{2}(\gamma(1+\beta\cos\theta),\sin\theta\cos\hat{\phi},-\sin\theta\sin\hat{\phi},-\gamma(\cos\theta+\beta)).
\end{eqnarray}
 {We calculate the
helicity amplitudes in this specific $t\bar{t}$ rest frame where both the Higgs and
the $e^+e^-$ system have common momenta in the negative $z$-axis as in
Eq.~(\ref{eq:ph}) and (\ref{eq:pee}), and the azimuthal angle
$\hat{\phi}$ between the $t\bar{t}$ plane and the $e\bar{e}$ plane are
given to the $e$ and $\bar{e}$ momenta.} 

We first note that the helicity amplitudes in Eq.~(\ref{eq10:amp}) can be factorized as 
\begin{eqnarray}\label{eq:amp}
{\cal M}_{\alpha\sigma\bar{\sigma}}
&=&\sum_{V=\gamma,Z}D_V(q)L^V_\mu(e\bar{e};\alpha)(-g^{\mu\nu})\hat{M}^V_\nu(ht\bar{t},\sigma,\bar{\sigma})\nonumber\\
&=&\sum_{V=\gamma,Z}D_V(q)L^V_\mu(e\bar{e};\alpha)(\sum_\lambda\epsilon^\mu(q,\lambda)^\ast\epsilon^\nu(q,\lambda))\hat{M}^V_\nu(ht\bar{t},\sigma,\bar{\sigma})\nonumber\\
&=&\sum_{V=\gamma,Z}D_V(q)\sum_{\lambda}L_{\lambda\alpha}^V(e\bar{e})\hat{M}^V_{\lambda\sigma\bar{\sigma}}(ht\bar{t})
\end{eqnarray}
where $q^\mu=p_e^\mu+p_{\bar{e}}^\mu$,
 $\sqrt{s}$ is the center-of-mass energy,
 $D_V(q)=1/(q^2-m_V^2+im_V^{}\Gamma_V^{})$ is {the Breit-Wigner
 propagator factor for $\gamma$, $Z$, and also $W$ for later use.} In Eq.~(\ref{eq:amp}), the $\gamma$ and $Z$ polarization factor $-g^{\mu\nu}$ is replaced by the sum 
 \begin{eqnarray}
\sum_{\lambda}\epsilon^{\mu}(q,\lambda)^\ast\epsilon^\nu(q,\lambda)=-g^{\mu\nu}+\frac{q^\mu q^\nu}{q^2}
\label{eq:eps}
\end{eqnarray}
with
\begin{eqnarray}
\epsilon^\mu(q,\lambda)=\begin{cases}
\frac{1}{\sqrt{2}} (0,\pm1,-i,0) \quad{\rm for} ~\lambda=\pm1\\
\gamma(\beta,0,0,-1)\quad~~ {\rm~~ for}~ \lambda=0
\end{cases}
\end{eqnarray}
along the $e\bar{e}$ four momentum
$
q^\mu=p_{e\bar{e}}^\mu=\sqrt{s}\gamma(1,0,0,-\beta)
$,
see Eq.~(\ref{eq:pee}),
because of the $e\bar{e}$ current conservation,
\begin{eqnarray}
q^\mu L_{\mu}^V(e\bar{e};\alpha)=0
\end{eqnarray}
in the $m_e=0$ limit. The leptonic currents are simply 
\begin{eqnarray}
L_\mu^V(e\bar{e};\alpha)=g_\alpha^{Vee}v^\dagger_{-\alpha}(\bar
 e)\sigma_\alpha^\mu u_\alpha(e)
\end{eqnarray}
where $u_\alpha$ and $v_\alpha$ are two-component chiral spinors with $\alpha=-1$ for $L$ ($\alpha=+1$ for $R$), and $\sigma_{\pm}^\mu=(1,\pm\vec{\sigma})$ are the chiral four-vectors of $\sigma$ matrices. The gauge couplings are 
\begin{eqnarray}
&&g_-^{Zf\bar{f}}=g_z(T_f^3-Q_f\sin^2\theta_W^{}), ~~g_+^{Zf\bar{f}}=-g_zQ_f\sin^2\theta_W^{}\nonumber\\
&&g_-^{\gamma f\bar{f}}=g_+^{\gamma f\bar{f}}=eQ_f=g_V^{\gamma f\bar{f}}
\end{eqnarray}
with $g_z=g/\cos\theta_W$ .
We find
\begin{eqnarray}
L_\mu^V(e\bar{e};\alpha)=\alpha g_\alpha^{Vee}\sqrt{2s}\epsilon_\mu (\vec{n},\alpha)
\end{eqnarray}
where the polarization direction $\vec{n}$ is along the $e$
three-momentum in the $e^+e^-$ rest frame, which is obtained from
Eq.~(\ref{eq:eps}) by a Lorentz boost with $\beta$ and $\gamma$ given in Eq.~(\ref{eq:betagamma}),
\begin{eqnarray}
\vec{n}=(-\sin\theta\cos\hat{\phi},\sin\theta\sin\hat{\phi},\cos\theta).
\label{eq:n}
\end{eqnarray}
The leptonic amplitudes are hence {expressed} explicitly as 
\begin{eqnarray}
{L_{\lambda\alpha}^V(e\bar{e}) = L_\mu^V(e\bar{e};\alpha)\eps^{\mu}(\vec{q},\lambda)^\ast=\alpha g_\alpha^{Vee}\sqrt{2s}\epsilon_\mu(\vec{n},\alpha){\eps^{\mu}(\vec{q},\lambda)^\ast}}
\end{eqnarray}
in terms of Wigner's $D$ functions in the $e\bar{e}$ rest frame, 
{where $\alpha$ denotes helicities along $\vec{n}$ in
Eq.~(\ref{eq:n}), while $\lambda$ denotes those along $\vec{q}$ ($-z$)
direction.} Summing up we find 
\begin{eqnarray}
{\cal M}_{\alpha\sigma\bar{\sigma}}
=
\sum_{V=\gamma,Z}R_{\alpha}^V\left[\frac{1-\alpha\cos\theta}{2}e^{-i\hat{\phi}}\hat{M}^V_{+\sigma\bar{\sigma}}+\frac{1+\alpha\cos\theta}{2}e^{i\hat{\phi}}\hat{M}^V_{-\sigma\bar{\sigma}}+\frac{\alpha\sin\theta}{\sqrt{2}}\hat{M}^V_{0\sigma\bar{\sigma}}\right],
\label{eq:amp_ass}
\end{eqnarray}
with $R_\alpha^V=-\alpha g_\alpha^{Vee}\sqrt{2s}D_V(q)$, where the three $D$ functions are shown explicitly.

In this expression Eq.~(\ref{eq:amp_ass}), the leptonic amplitudes $L_{\lambda\alpha}^V(e\bar{e})$ gives the kinematical dependence on the production scattering angle $\theta$ and the azimuthal angle $\hat{\phi}$ in terms of Wigner's $D$ functions, whereas the $ht\bar{t}$ production amplitudes $\hat{M}^V_{\alpha\sigma\bar{\sigma}}$ depend only on the $t\bar{t}$ invariant mass $m_{t\bar{t}}$, and the polar angle of the top quark momentum $\hat{\theta}$.
After using the equations of motion for $t$ and $\bar{t}$, we find a compact expression for $\hat{M}_{\alpha\sigma\bar{\sigma}}^V$,
\begin{eqnarray}
\hat{{\cal M}}_{\lambda\sigma\bar{\sigma}}^V&=&\epsilon(q,\lambda)_\nu\Bigg\{2m_tg_{htt}\cos\xi_{htt}^{}(D_t^1+D_t^2)\left[g_-^{Vtt}u_L^\dagger\sigma_-^\nu\nu_L^{}+g_+^{Vtt}u_R^\dagger\sigma_+^\nu v_R^{}\right]\nonumber\\
&+&g_{hZZ}^{}\delta_{VZ}D_Z^{Q}\left[g_-^{Ztt}u_L^\dagger\sigma_-^\nu\nu_L+g_+^{Ztt}u_R^\dagger\sigma_+^\nu v_R^{}+\frac{mQ^\nu}{m_Z^2}g_A^{Ztt}(u_L^{\dagger}v_R-u_R^+v_L)\right]\nonumber\\
&+&g_{htt}^{}D_t^1\left[e^{-i\xi_{htt}^{}}g_-^{Vtt}u_R^\dagger\sigma_+\cdot p_h\sigma_-^\nu v_L+e^{i\xi_{htt}^{}}g_+^{Vtt}u_L^\dagger\sigma_-\cdot p_h \sigma_+^\nu v_R \right] \nonumber\\
&-&g_{htt}^{}D_t^2\left[e^{i\xi_{htt}}g_-^{Vtt}u_L^{\dagger}\sigma_-^\nu \sigma_+\cdot p_h v_R + e^{-i\xi_{htt}}g_+^{Vtt} u_R^{\dagger}\sigma_+^\nu \sigma_-\cdot p_h v_L \right]\Bigg\},
\end{eqnarray}
where
 \begin{eqnarray}
 D_t^1=D_t(q-p_t),~~~~D_t^2=D_t(p_t-q),~~~~D_Z^Q=D_Z(Q)~~~ {\rm with}~~~Q=p_t+p_{\bar{t}},
\end{eqnarray}
are the $t$ and $Z$ Breit-Wigner propagator factors.

For $\lambda=\pm1$, the matrix elements are proportional to the Wigner's $D$
functions in the $t\bar{t}$ rest frame:
\begin{eqnarray}
\hat{\cal M}_{\lambda\pm\mp}^{V}&=&\frac{\pm1-\lambda\cos\hat{\theta}}{\sqrt{2}}
\Bigg\{
\Bigg[4m_tg_{htt}\cos\xi_{htt}^{} (D_{t}^1+D_{t}^2)+2g_{hZZ}^{}\delta_{VZ}^{}D_Z^{Q}\Bigg]\left(g^V_{\,V}\hat{E}\mp g^V_{\,A} \hat{p}\right)\nonumber\\
&+&2m_t^{}g_{htt}\cos\xi_{htt}^{}\left[g^V_{\,V}(D_t^1+D_t^2)\left(\sqrt{s}\gamma-m_{t\bar{t}})-g^V_{\,A}(D_t^1-D_t^2)\lambda \sqrt{s}\gamma\beta\right)\right]\nonumber\\
&+&2im_{t}^{}g_{htt}\sin\xi_{htt}^{}\left[g^V_{\,V}(D_t^1+D_t^2) \lambda \sqrt{s}\gamma\beta- g^V_{\,A}(D_t^1-D_t^2)(\sqrt{s}\gamma-m_{t\bar{t}})\right]
\Bigg\}
\label{eq:lampmmp}
\end{eqnarray}
for $(\sigma-\bar{\sigma})/2=\pm1$, and
\begin{eqnarray}
\hat{M}_{\lambda\pm\pm}^{V}&=&\mp\lambda \frac{\sin\hat{\theta}}{\sqrt{2}}
\Bigg\{2m_tg^V_{\,V}
\Bigg[2m_tg_{htt}\cos\xi_{htt} (D_{t}^1+D_{t}^2)+g_{hZZ}^{}\delta_{VZ}^{}D_Z^{Q}\Bigg]\nonumber\\
&+&2g_{htt}\cos\xi_{htt}\left[g^V_{\,V}(D_t^1+D_t^2)\left(\hat{E}(\sqrt{s}\gamma-m_{t\bar{t}})\mp\lambda
						     \hat{p}\sqrt{s}\gamma\beta\right)\pm
g^V_{\,A}(D_t^1-D_t^2)(\hat{p}(\sqrt{s}\gamma-m_{t\bar{t}})\mp\lambda\hat{E}\sqrt{s}\gamma\beta)\right]\nonumber\\
&\mp&2ig_{htt}\sin\xi_{htt}\left[g^V_{\,V}(D_t^1+D_t^2)\left(\hat{p}(\sqrt{s}\gamma-m_{t\bar{t}})\mp\lambda \hat{E}\sqrt{s}\gamma\beta\right)\pm g^V_{\,A}(D_t^1-D_t^2)(\hat{E}(\sqrt{s}\gamma-m_{t\bar{t}})\mp\lambda \hat{p}\sqrt{s}{\gamma\beta})\right]\Bigg\}\nonumber
\label{eq:lampmpm}\\
\end{eqnarray}
for $(\sigma-\bar{\sigma})/2=0$.
Here we introduce compact notation,
\begin{eqnarray}
{
g^V_{\,V}=(g_-^{Vtt}+g_+^{Vtt})/2,\quad
g^V_{\,A}=(g_-^{Vtt}-g_+^{Vtt})/2},
\end{eqnarray}
 where 
\begin{eqnarray}
\hat{E}=m_{t\bar{t}}/2 \quad{\rm and}\quad
\hat{p}=\hat{E}\hat{\beta}
\end{eqnarray}
are the top energy and momentum in the
$t\bar{t}$ rest frame.

For $\lambda=0$, the amplitudes are still proportional to the
Wigner's $D$ function when $(\sigma-\bar{\sigma})/2=\pm1$, whereas they
are the sum of the terms for the $J=1$ component and a constant term for
the $J=0$ component when $(\sigma-\bar{\sigma})/2=0$:
\begin{eqnarray}
\hat{\cal M}_{0\pm\mp}^{V}&=&-\sin\hat{\theta}\Bigg\{2\gamma\left[2m_tg_{htt}^{}\cos\xi_{htt}(D_t^1+D_t^2)+g_{hZZ}\delta_{VZ}D_Z^Q\right]\left(\hat{E}g^V_{\,V}\mp \hat{p}g^V_{\,A}\right)\nonumber\\
&+&2g_{htt}m_t\hat{E}_h\left[\cos\xi_{htt}g^V_{\,V}(D_t^1+D_t^2)-i\sin\xi_{htt}g^V_{\,A}(D_t^1-D_t^2)\right]
\Bigg\}
\label{eq:0pmmp}
\\
\hat{\cal M}_{0\pm\pm}^V&=&4m_t^2g_{htt}\cos\xi_{htt} (D_t^1+D_t^2)\gamma \left(\pm g^V_{\,V}\cos\hat{\theta}-g^V_{\,A}\beta\right)\nonumber\\
&+&2m_tg_{hZZ}\delta_{VZ}D_Z^Q\left[\gamma(\pm g^V_{\,V}\cos\hat{\theta}-g^V_{\,A}\beta)+\frac{mp_h}{m_Z^2}g^V_{\,A}\right]\nonumber\\
&+&2g_{htt}^{}\cos\xi_{htt}\left[\hat{E}\left(\pm E_h\cos\hat{\theta} g^V_{\,V}+p_h g^V_{\,A}\right)(D_t^1+D_t^2)\pm \hat{p}(\pm E_h\cos\hat{\theta} g^V_{\,A}+p_h g^V_{\,V})(D_t^1-D_t^2)\right]\nonumber\\
&-&2ig_{htt}^{}\sin\xi_{htt}\left[\pm \hat{p}(\pm E_h\cos\hat{\theta} g^V_{\,V}+p_h g^V_{\,A})(D_t^1+D_t^2)+\hat{E}(\pm E_h\cos\hat{\theta} g^V_{\,A}+p_h g^V_{\,V})(D_t^1-D_t^2)\right]
\label{eq:0pmpm}
\end{eqnarray}
Here, $E_h=\sqrt{s}-E_{t\bar{t}}=\sqrt{s}-m_{t\bar{t}}\gamma$ and $p_h=p_{t\bar{t}}=m_{t\bar{t}}\gamma\beta$ are the Higgs energy and the momentum in the $e\bar{e}$ rest frame.
We note here that the terms proportional to $\sin\xi$ in all the 12
amplitudes
Eqs.~(\ref{eq:lampmmp},\ref{eq:lampmpm},\ref{eq:0pmmp},\ref{eq:0pmpm}) are
either proportional to $\hat{p}=\hat{E}\hat{\beta}$, $\beta$, or
$D_t^1-D_t^2$, which are all {suppressed} strongly near
the $ht\bar{t}$ production threshold.

\begin{figure}[t]
\includegraphics[width=0.6\textwidth]{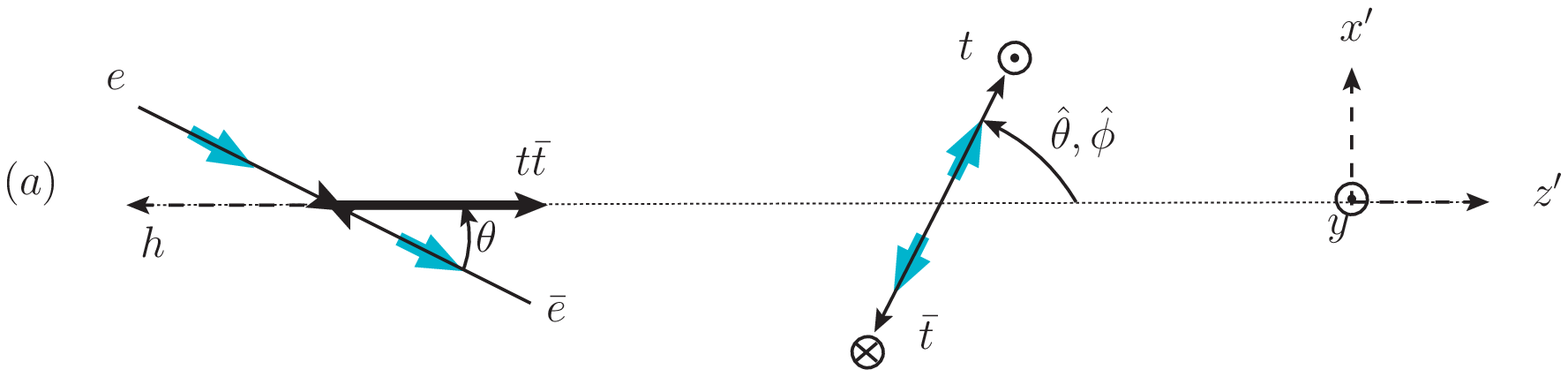}\\
\includegraphics[width=0.6\textwidth]{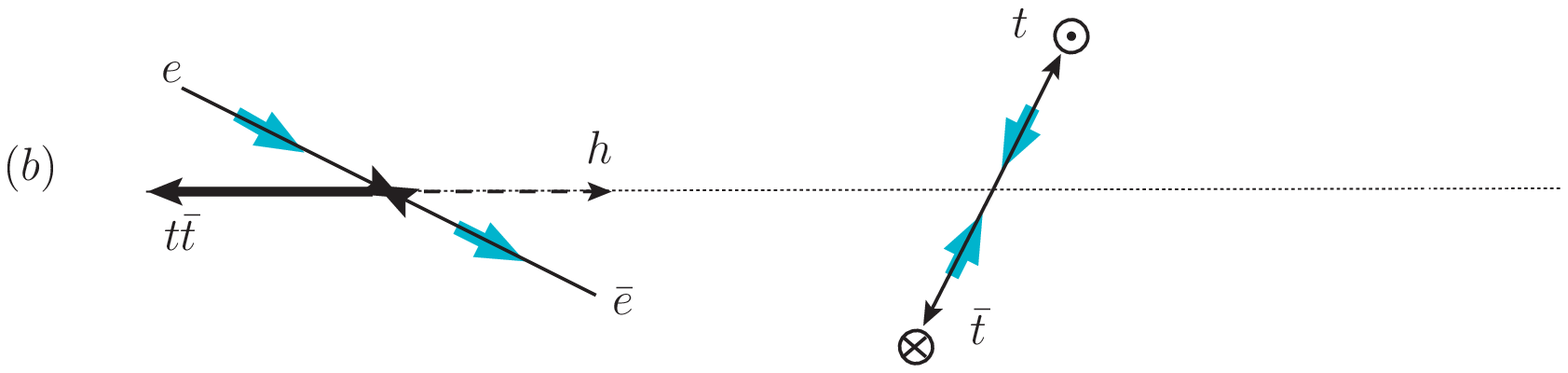}\\
\includegraphics[width=0.6\textwidth]{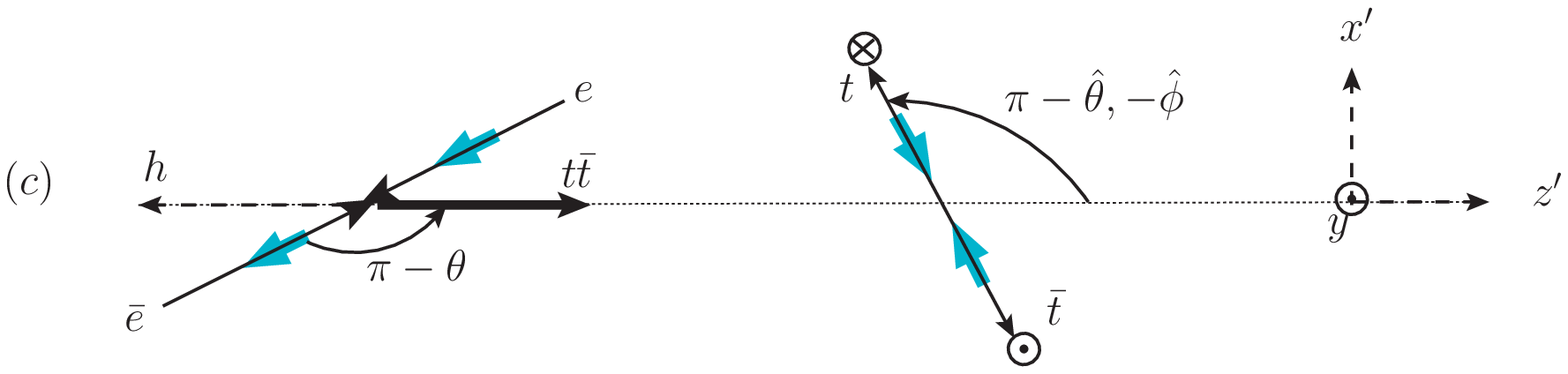}
\caption{CP transformation of $e\bar{e}\to ht\bar{t}$ helicity amplitudes. Green shaded arrows show the spin direction along the particle momenta. The $e$ and $\bar{e}$ momenta are shown in the $e\bar{e}$ rest frame, whereas $t$ and $\bar{t}$ momenta are shown in the $t\bar{t}$ rest frame. The common $z^\prime$-axis is chosen along the $t\bar{t}$ momentum in the $e\bar{e}$ rest frame, or the negative of the $e\bar{e}$ and the $h$ momentum in the $t\bar{t}$ rest frame, and the common $y$-axis is vertical to the $e\bar{e}\to h+(t\bar{t})$ scattering plane; $\vec{{}n}_y=\frac{\vec{{}p}_e\times\vec{{}p}_{t\bar{t}}}{|\vec{{}p}_e\times\vec{{}p}_{t\bar{t}}|}$ in the $e\bar{e}$ rest frame. (a) shows the momenta and spins when the helicities of $e,~t~{\rm and}~\bar{t}$ are all $+\frac{1}{2}$, ($\alpha=\sigma=\bar{\sigma}=+$). The symbol $\odot~(\otimes)$ shows that the $y$-component of each momentum is positive (negative). Hence $0<\hat{\theta},\hat{\phi}<\pi/2$. (b) shows the process with ${\rm CP}$ transformed particles and their vectors in the original frames. (c) is obtained from (b) by two rotations, rotations of $\pi$ about the $z^\prime$-axis and by the $y$-axis. The coordinate system parameterizing the three momentum is the same as in (a), and we can read off the ${\rm CP}$ transformation of the helicity amplitudes as given by Eq.~(\ref{eq:CP_amp}). 
 }\label{fig:CP}
\end{figure}

It is instructive to note here that the above amplitudes satisfy the CP transformation properties
\begin{eqnarray}
 M_{\alpha,\sigma,\bar{\sigma}}(\theta,\hat{\theta},{\hat\phi};\xi)=M_{\alpha,-\bar{\sigma},-\sigma}(\pi-\theta,\pi-\hat{\theta},{-\hat\phi};-\xi)
 \label{eq:CP_amp}
 \end{eqnarray}
which is illustrated in Fig.~\ref{fig:CP}. 
The upper plot (a) shows the three momenta and the helicities of $e$ and $\bar{e}$ in the $e\bar{e}$ rest frame, as well as those of $t$ and $\bar{t}$ in the $t\bar{t}$ rest frame.
The common $z^\prime$-axis is chosen along the $t\bar{t}$ momentum direction in the $e\bar{e}$ rest frame. The electron momentum is in the $x^\prime$-$z^\prime$ plane.
  The vertical arrows ($\odot$ and $\otimes$) show that the $t$-momentum has positive $y$-component, or $0<\hat{\theta}$, $\hat{\phi}<\pi/2$. 
Shaded arrows show all the fermion spin directions for $\alpha=\sigma=\bar{\sigma}=+1$. The middle plot (b) is obtained from (a) by the CP transformation, which reverses the sign of all three momenta, but keeps the spin (hence the helicity is reserved), and exchange particles and antiparticles. Note that the initial $e\bar{e}$ state is invariant under CP transformation for each helicity. The momentum configuration of (b) is expressed in our reference frame as in the bottom plot (c), in which $\vec{p}_{t\bar{t}}$ is along the $z^\prime$ axis and the $y$ axis is along $\vec{p}_e\times\vec{p}_{t\bar{t}}$. The frame (c) is obtained from (b) by rotations, which do not affect the helicity amplitudes. This gives the identity Eq.~(\ref{eq:CP_amp}), which show explicitly how CP-violating term proportional to $\sin\xi_{htt}$ should behave under exchange of angular variables:
\begin{eqnarray}
(\theta,\hat{\theta},\hat{\phi})\leftrightarrow(\pi-\theta,\pi-\hat{\theta},-\hat{\phi}),
\end{eqnarray}
and when the $t$ and $\bar{t}$ helicities are exchanged as 
\begin{eqnarray}
(\sigma,\bar{\sigma})\leftrightarrow(-\bar{\sigma},-\sigma).
\end{eqnarray}

The helicity amplitudes $\hat{M}^V_{\alpha\sigma\bar{\sigma}}$ contain all the information about the $ht\bar{t}$ (and the $hZZ$) coupling that we can probe in the process $e^+e^-\to ht\bar{t}$. Being complex numbers, however, they are not directly observable in experiments. For instance, the differential cross section 
\begin{eqnarray}
d\sigma=\frac{1}{2s}\frac{1}{4}\sum_\alpha\sum_\sigma\sum_{\bar{\sigma}}|M_{\alpha\sigma\bar{\sigma}}|^2d\Phi_{htt}
\end{eqnarray}
with the 3-body phase space
\begin{eqnarray}
d\Phi_{htt}={\frac{1}{64\pi^2}}\bar{\beta}\hat{\beta}\frac{dm_{t\bar{t}}^2}{2\pi}\frac{d\cos\theta d\cos\hat{\theta}d\hat{\phi}}{8\pi}
\end{eqnarray}
with $\bar{\beta}$ in Eq.~(\ref{eq:betabar}) and $\hat{\beta}$ in Eq.~(\ref{eq:betahat}), measures only the {squared} sum of all the helicity amplitudes. With $e^-$ (and possibly $e^+$) beam polarization, {the sum of} $e_L\bar{e}_R$ annihilation ($\alpha=-1$) and that of $e_R\bar{e}_L$ annihilation ($\alpha=+1$) can be resolved.

When we study $t$ and $\bar{t}$ decay distributions and their
correlations, we can measure 15 more combinations of the helicity
amplitudes (16 each, including the absolute value squared of the 4
amplitudes for $\alpha=-1$ and $\alpha=+1$ with beam polarization). In this section, we illustrate this when both $t$ and $\bar{t}$ decay semi-leptonically,
\begin{eqnarray}
e^+e^-\to ht\bar{t}\to h(b\ell\bar{\nu}_\ell)(\bar{b}\ell^\prime\bar{\nu}_{\ell^\prime})
\label{eq:eehtt}
\end{eqnarray}
The helicity amplitude for the process can be expressed (for $e_\alpha\bar{e}_{-\alpha}$ collisions) as 
\begin{eqnarray}
{\cal M}_{\alpha}=\sum_\sigma\sum_{\bar{\sigma}}M_{\alpha\sigma\bar{\sigma}}D_t(p_t)D_{{t}}(p_{\bar{t}})M_\sigma\bar{M}_{\bar{\sigma}}
\end{eqnarray}
with the Breit-Wigner propagator factors
\begin{eqnarray}
D_t(p)={\frac{1}{p^2-m_t^2+im_t\Gamma_t}}
\end{eqnarray}
and the decay amplitudes
\begin{subequations}
\begin{align}
M_\sigma&=\frac{g^2}{2}D_W^{}(p_t-p_b)~
u_L^\dagger(p_b)\sigma_-^\mu u_L(p_t,\sigma)~
u_L^\dagger(p_\nu)\sigma_{-\mu}v_L(p_{\bar{\ell}})\label{eq:decaymatr}\\
\bar{M}_{\bar{\sigma}}&=\frac{g^2}{2}D_W^{}(p_{\bar{t}}-p_{\bar{b}})~
v_L^\dagger(p_{\bar{t}},\bar{\sigma})\sigma_-^\mu v_L(p_{\bar{b}})
~u_L^\dagger(p_\ell)\sigma_{-\mu}v_L(p_{\bar{\nu}})\label{eq:decaymatr-anti}
\end{align}
\end{subequations}
The differential cross section for the process Eq.~(\ref{eq:eehtt}) is hence (for unpolarized beams) 
\begin{eqnarray}
d\sigma=\frac{1}{2s}\frac{1}{4}\sum_\alpha|{\cal M}_\alpha|^2d\Phi_7
\label{eq:dsigma}
\end{eqnarray}
where the {7-body} phase space can be decomposed as 
\begin{eqnarray}
d\Phi_7=d\Phi_{htt}(m_h^2,p_t^2,p_{\bar{t}}^2)\frac{dp_t^2}{2\pi}\frac{dp_{\bar{t}}^2}{2\pi}d\Phi_{b\bar{\ell}\nu_\ell}(p_t^2)d\Phi_{\bar{b}\ell^\prime\bar{\nu}_{\ell^\prime}}(p_{\bar{t}}^2)
\end{eqnarray}
{with $p_t^2$ and $p_{\bar{t}}^2$ as the invariant mass squared} of the ($b\bar{\ell}\nu_{\bar{\ell}}$) and ($\bar{b}\ell^\prime\bar{\nu}_{\bar{\ell}^\prime}$) systems, respectively. In the narrow width limit of the top quark,
\begin{eqnarray}
\int \frac{dp_t^2}{2\pi}|D_t(p_t)|^2=\int \frac{d p_{\bar{t}}^2}{2\pi}|D_{{t}}({p_{\bar{t}}})|^2=\frac{1}{2m_t\Gamma_t}
\end{eqnarray}
holds and the differential cross section in Eq.~(\ref{eq:dsigma}) can be expressed as 
\begin{eqnarray}
d\sigma=\frac{1}{2s}\frac{1}{4}\sum_\alpha\left|\sum_\sigma\sum_{\bar{\sigma}}M_{\alpha\sigma\bar{\sigma}}M_\sigma\bar{M}_{\bar{\sigma}}\right|^2 
d\Phi_{ht\bar{t}}~
\frac{d\Phi_{b\bar{\ell}\nu_\ell}}{2m_t\Gamma_t}~
\frac{d\Phi_{\bar{b}\ell^\prime\bar{\nu}_{\ell^\prime}}}{2m_t\Gamma_t}.
\end{eqnarray}
The above expression can be expressed as 
\begin{eqnarray}\label{eq:diff_xs_set}
d\sigma=\sum_\sigma\sum_{\bar{\sigma}}\sum_{\sigma^\prime}\sum_{\bar{\sigma^\prime}}
&&\left(\frac{1}{2s}\frac{1}{4}\sum_\alpha(M_{\alpha\sigma\bar{\sigma}})(M_{\alpha\sigma^\prime\bar{\sigma}^\prime})^\ast d\Phi_{ht\bar{t}}\right)d\rho_{\sigma\sigma^\prime}d\bar{\rho}_{\bar{\sigma}\bar{\sigma}^\prime}
\label{eq:dxsrho}
\end{eqnarray}
with
\begin{subequations}
\begin{align}
d\rho_{\sigma\sigma^\prime}&=M_{\sigma}M_{\sigma^\prime}^\ast d\Phi_{b\bar{\ell}\nu_\ell}/(2m_t\Gamma_t),
\label{eq:drhoss}\\
d\bar{\rho}_{\bar{\sigma}\bar{\sigma}^\prime}&=M_{\bar{\sigma}}M_{\bar{\sigma}^\prime}^\ast d\Phi_{{\bar{b}\ell^\prime\bar{\nu}_{\ell^\prime}}}/(2m_t\Gamma_t).
\label{eq:drhobss}
\end{align}
\label{eq:drhoss&bss}
\end{subequations}
\hspace*{-0.16cm}The differential decay density matrices in Eqs.~(\ref{eq:drhoss&bss}) are calculated in Appendix ~\ref{sec:appA}, and take particularly simple form for semi-leptonic decay 
\begin{subequations}
\begin{align}
d\rho&=
\left(\begin{array}{cc}
d\rho_{++} &d\rho_{+-}\\
d\rho_{-+}&d\rho_{--}
\end{array}\right)
=
\left(\begin{array}{cc}
1+\cos\bar{\theta}^\ast&\sin\bar{\theta}^\ast e^{i\bar{\phi}^\ast}\\
\sin\bar{\theta}^\ast e^{-i\bar{\phi}^\ast}&1-\cos\bar{\theta}^\ast
\end{array}\right)
B_\ell
\frac{d\cos\bar{\theta}^\ast d\bar{\phi}^\ast}{4\pi}
\label{eq:drho}\\
d\bar{\rho}&=
\left(\begin{array}{cc}
d\bar{\rho}_{++} &d\bar{\rho}_{+-}\\
d\bar{\rho}_{-+}&d\bar{\rho}_{--}
\end{array}\right)
=
{
\left(\begin{array}{cc}
{1+\cos{\theta}^\ast}&{\sin{\theta}^\ast e^{-i{\phi}^\ast}}\\
{\sin{\theta}^\ast e^{i{\phi}^\ast}}&{1-\cos{\theta}^\ast}
\end{array}\right)
}
B_\ell
\frac{d\cos\theta^\ast d\phi^\ast}{4\pi},\
\label{eq:drhob}
\end{align}\label{eq:drho2}
\end{subequations}  
\hspace*{-0.21cm}
where $B_\ell=\sum_\ell B(t\to b\bar{\ell}\nu_\ell)=\sum_\ell B(\bar{t}\to\bar{b}\ell\bar{\nu}_\ell)\simeq0.33$ is the semi-leptonic branching fractions summed over $\ell=e,\mu,\tau$. 
Here, $\bar{\theta}^\ast$ and $\bar{\phi}^\ast$ ($\theta^\ast$ and $\phi^\ast$) are the polar and azimuthal angles of $\bar{\ell}~(\ell)$ in the $t~(\bar{t})$ rest frame where the polar axis is chosen along the $t$ momentum direction in the $t\bar{t}$ rest frame. More details are explained in Appendix~\ref{sec:appA}.
By inserting Eq.~(\ref{eq:drho2}) into Eq.~(\ref{eq:dxsrho}), we find 
\begin{eqnarray}
&&\frac{d\sigma_\alpha}{d\Phi_{htt}^{}d\cos\bar{\theta}^\ast d\bar{\phi}^\ast d\cos\theta^\ast d\phi^\ast}\nonumber\\
&=&\frac{B_\ell^2}{(4\pi)^2}\times\Big\{\nonumber\\
&&
\quad|M_{\alpha++}|^2(1+\cos\bar{\theta}^\ast){(1+\cos{\theta}^\ast)}+|M_{\alpha+-}|^2(1+\cos\bar{\theta}^\ast){(1-\cos{\theta}^\ast)}\nonumber\\
&&+|M_{\alpha--}|^2(1-\cos\bar{\theta}^\ast){(1-\cos{\theta}^\ast)}+|M_{\alpha-+}|^2(1-\cos\bar{\theta}^\ast){(1+\cos{\theta}^\ast)}\nonumber\\[2mm]
&&{+}2\left[{\rm Re}(M_{\alpha++}M_{\alpha+-}^\ast)\cos\phi^\ast{+}{\rm Im}(M_{\alpha++}M_{\alpha+-}^\ast)\sin\phi^\ast\right]\sin\theta^\ast(1+\cos\bar{\theta}^\ast)\nonumber\\
&&{+}2\left[{\rm Re}(M_{\alpha-+}M_{\alpha--}^\ast)\cos\phi^\ast{+}{\rm Im}(M_{\alpha-+}M_{\alpha--}^\ast)\sin\phi^\ast\right]\sin\theta^\ast(1-\cos\bar{\theta}^\ast)\nonumber\\
&&+2\left[{\rm Re}(M_{\alpha++}M_{\alpha-+}^\ast)\cos\bar{\phi}^\ast-{\rm Im}(M_{\alpha++}M_{\alpha-+}^\ast)\sin\bar{\phi}^\ast\right]\sin\bar{\theta}^\ast{(1+\cos\theta^\ast)}\nonumber\\
&&+2\left[{\rm Re}(M_{\alpha+-}M_{\alpha--}^\ast)\cos\bar{\phi}^\ast-{\rm Im}(M_{\alpha+-}M_{\alpha--}^\ast)\sin\bar{\phi}^\ast\right]\sin\bar{\theta}^\ast{(1-\cos\theta^\ast)}\nonumber\\[2mm]
&&{+}2\left[{\rm Re}(M_{\alpha++}M_{\alpha--}^\ast)\cos{(\bar{\phi}^\ast-\phi^\ast)}-{\rm Im}(M_{\alpha++}M_{\alpha--}^\ast)\sin{(\bar{\phi}^\ast-\phi^\ast)}\right]\sin\bar{\theta}^\ast\sin\theta^\ast\nonumber\\
&&{+}2\left[{\rm Re}(M_{\alpha+-}M_{\alpha-+}^\ast)\cos{(\bar{\phi}^\ast+\phi^\ast)}-{\rm Im}(M_{\alpha+-}M_{\alpha-+}^\ast)\sin{(\bar{\phi}^\ast+\phi^\ast)}\right]\sin\bar{\theta}^\ast\sin\theta^\ast\Big\}.
\label{eq:16term}
\end{eqnarray}
\begin{figure}[h!]
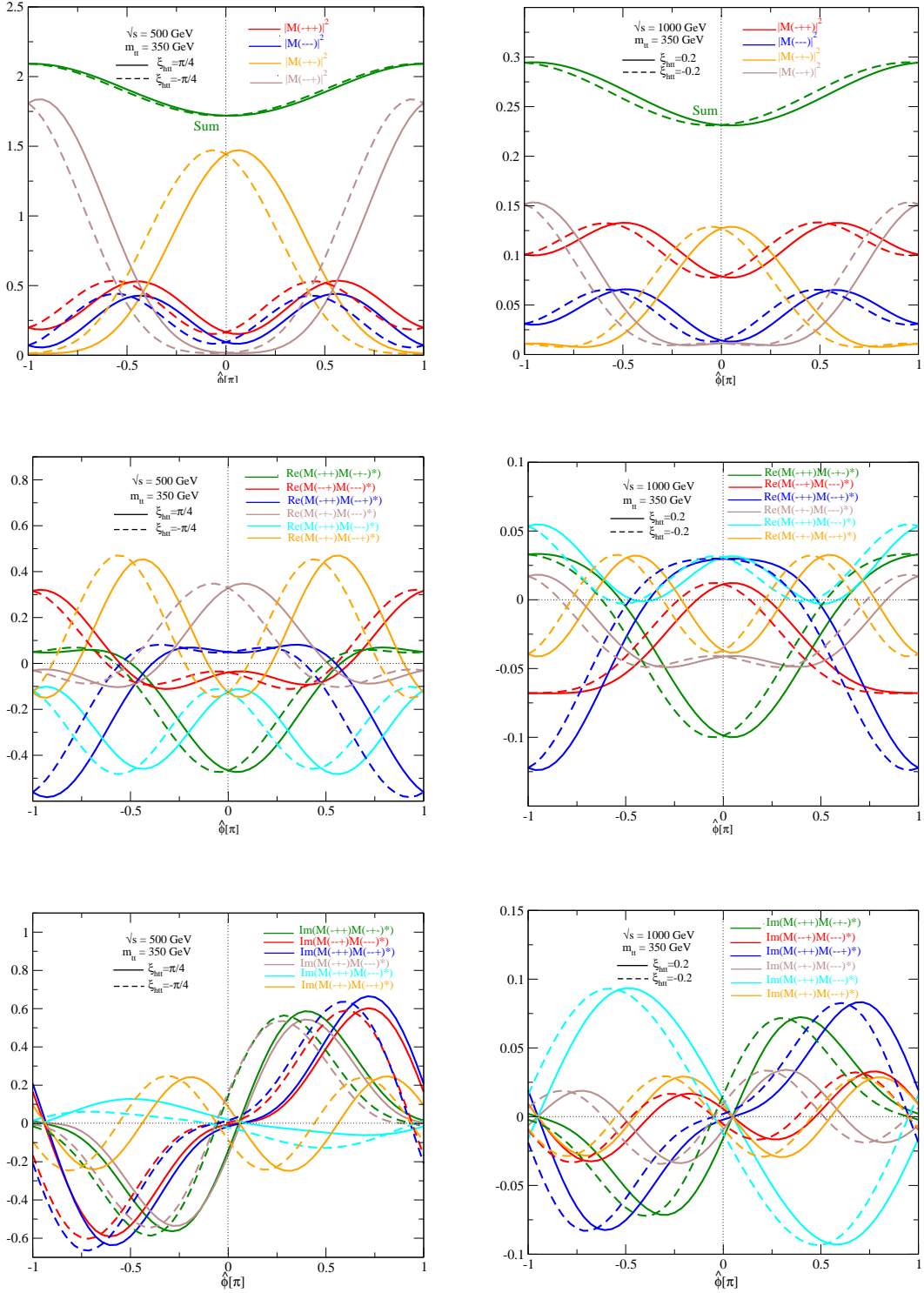

\vspace{0.5cm}
%\begin{centering}
%\begin{tabular}{c}
\includegraphics[width=0.36\textwidth]{500M4_eL.eps}
\hspace{0.8cm}
\includegraphics[width=0.365\textwidth]{1000M4_eL.eps}\\
\vspace{1cm}
\includegraphics[width=0.36\textwidth]{500Re_eL.eps}
\hspace{0.8cm}
\includegraphics[width=0.365\textwidth]{1000Re_eL.eps}\\
\vspace{1cm}
\includegraphics[width=0.36\textwidth]{500Im_eL.eps}
\hspace{0.8cm}
\includegraphics[width=0.365\textwidth]{1000Im_eL.eps}\\
%\vspace{0.1cm}
%\end{tabular}
\caption {The azimuthal angle distribution of the 16 helicity amplitudes
 combinations $
 M_{\alpha\sigma\bar{\sigma}}^{}M_{\alpha\sigma^\prime\bar{\sigma}^\prime}^\ast$
 for $\alpha=-1~(e_L^{}\bar{e}_R^{}$ annihilation) at $\sqrt{s}=500$~GeV
 for $\xi_{htt}^{}=\pm\pi/4$~(left panels) and at $\sqrt{s}=1000$~GeV
 for $\xi_{htt}=\pm0.2$~(right panels). We set $\cos\theta=0$,
 $\cos\hat{\theta}=0.5$ and $m_{t\bar{t}}=350$~GeV in all the plots. Solid
 curves are for $\xi_{htt}>0$, and dashed curves are for
 $\xi_{htt}<0$. }
\label{fig:azm}
%\end{centering}
\end{figure}
It is now clear that all the real and imaginary parts of the product of the helicity amplitudes 
$M_{\alpha\sigma\bar{\sigma}}$
 and its complex conjugates 
 $M_{\alpha\sigma^\prime\bar{\sigma}^\prime}^\ast$
  including $\sigma^\prime\neq\sigma$
   and $\bar{\sigma}^\prime\neq\bar{\sigma}$ 
   can be measured by studying the correlated decays $t\to
   b\bar{\ell}\nu$ and $\bar{t}\to\bar{b}\ell\bar{\nu}$ at all
   $ht\bar{t}$ phase space point $(m_{t\bar{t}},\cos\theta,\cos\hat{\theta},\hat{\phi})$. There are 
   {four $\sum_\alpha|M_{\alpha\sigma\bar{\sigma}}|^2$
    terms, six 
    $\sum_\alpha{\rm Re}(M_{\alpha\sigma^\prime\bar{\sigma}^\prime}^{}M_{\alpha\sigma^\prime\bar{\sigma}^\prime}^\ast)$ 
    terms and six 
    $\sum_\alpha{\rm Im}(M_{\alpha\sigma\bar{\sigma}}^{}M_{\alpha\sigma^\prime\bar{\sigma}^\prime}^\ast)$ terms.} 
    With polarized $e$ beams, $\alpha=-1$ and $\alpha=+1$ combinations can be resolved. 

In Fig.~\ref{fig:azm}, we show the $\hat{\phi}$ distribution of all {
the 16 combinations of
$M_{\alpha\sigma\bar{\sigma}}^{}M_{\alpha\sigma^\prime\bar{\sigma}^\prime}^\ast$ } for $\alpha=-1$ case ($e_L\bar{e}_R$ annihilation) at
$\sqrt{s}=500$~GeV~(left) and at $\sqrt{s}=1000$~GeV~(right). We set
$\cos\theta=0,\cos\hat{\theta}=0.5$ and $m_{t\bar{t}}=350$~GeV at both
energies where the $\xi_{htt}^{}$-dependences are found to be
significant. We compare $\xi_{htt}=\frac{\pi}{4}$~(solid lines) and
$\xi_{htt}=-\frac{\pi}{4}$~(dashed lines) for $\sqrt{s}=500$~GeV, and
$\xi_{htt}=\pm0.2$ for $\sqrt{s}=1000$~GeV.

The top panels show the {four} absolute value squared
$|M_{-\sigma\bar{\sigma}}|^2$ and their sum. CP violation appears as a
phase shift in the $\hat{\phi}$ distribution whose sign and magnitude
are proportional to $\xi_{ htt}^{}$~\cite{Hagiwara:2016rdv}. The
difference is reduced significantly when only the total sum of all
squared amplitudes, i.e.\ the $ht\bar{t}$ distributions are observed. The polar angle distributions of $t$ and $\bar{t}$ decays can resolve the {four} individual contributions $(\sigma\bar{\sigma})=(+,-),(-,+),(++),(--)$, according to Eq.~(\ref{eq:16term}).

The middle panels show the real part of the six off-diagonal
($\sigma^\prime\neq\sigma$ or $\bar{\sigma}^\prime\neq\bar{\sigma}$)
products.
{In the absence of CP violation, these are even functions of
$\hat\phi$. The CP-violating asymmetries appear again as a phase shift or asymmetries between $\hat{\phi}>0$ and $\hat{\phi}<0$.} From Eq.~(\ref{eq:16term}), we learn that the 6 real terms are measured as coefficients of $\cos\phi^\ast$, $\cos\bar{\phi}^\ast$, $\cos(\bar{\phi}^\ast+\phi^\ast)$ or $\cos(\bar{\phi}^\ast-\phi^\ast)$.

In the bottom panel we show the corresponding imaginary parts of the
{six} off-diagonal products. They are measured as coefficients of
$\sin\phi^\ast$, $\sin\bar{\phi}^\ast$,
$\sin(\bar{\phi}^\ast+\phi^\ast)$ or $\sin(\bar{\phi}^\ast-\phi^\ast)$
according to Eq.~(\ref{eq:16term}).
{It is worth noting that these distributions are odd
functions of $\hat\phi$ if there is no CP violation~($\xi_{htt}=0$).
CP violation induces a phase shift in these distributions whose
sign and magnitude are proportional to $\xi_{htt}$. 
}

We also study all the distributions for $\alpha=+1~(e_R\bar{e}_L$ annihilation), but they are found to be very similar to the $\alpha=-1$ case shown in Fig.~\ref{fig:azm}, with significantly smaller magnitudes. Although the results shown in Fig.~\ref{fig:azm} are for a particular kinematical configuration of $ht\bar{t}~ (m_{t\bar{t}}=350~{\rm GeV},\cos\theta=0,\cos\hat{\theta}=0.5)$, and for $\alpha=-1~(e_L\bar{e}_R$ annihilation), their dependence on the sign of $\xi_{htt}$ shows the possible improvement in the CP-violation discovery potential in $e^+e^-$ collision experiments, by making use of all the information given by $t$ and $\bar{t}$ decay angular correlations.
\section{Cross sections and QCD corrections}\label{sec:results}
In this section we study the energy dependence of the total cross sections and the QCD higher-order corrections, perturbative NLO corrections and resummation of Coulombic corrections that account for topponium formation below and around the threshold, $m_{t\bar{t}}\sim2m_t$. Those studies are made in the two CP-conserving limits, at $\xi_{htt}=0~(h=H$, the SM limit), and at $\xi_{htt}=\frac{\pi}{2}~(h=A$, the pseudo scalar limit). The results are used to normalize our statistical analysis in the next sections, which are based on the leading-order (LO) matrix elements.

\subsection{Leading-order production cross section}
\begin{figure}[t]
\begin{centering}
\begin{tabular}{c}
\includegraphics[width=0.45\textwidth]{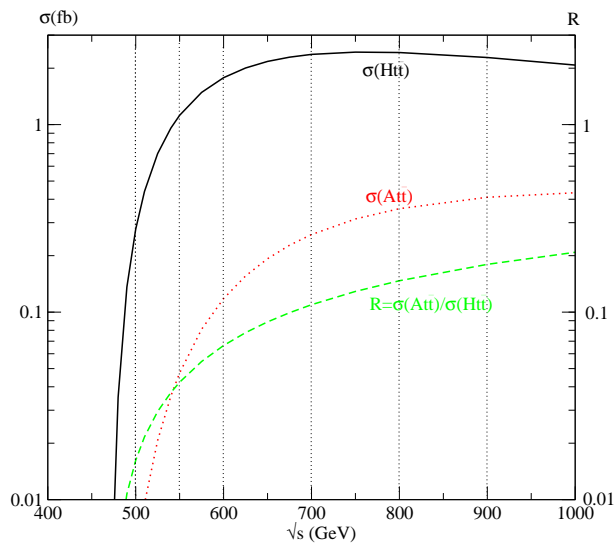}
\end{tabular}
\caption {The leading-order total cross section for $e^+e^-\to ht\bar{t}$ for the pure CP-even ($h=H$) and CP-odd ($h=A$) limits. $\sigma(Ht\bar{t})$
and $\sigma(At\bar{t})$ are shown by black solid and red dotted curves respectively, and their ratio $R=\sigma(At\bar{t})/\sigma(Ht\bar{t})$ is given by green dashed line.    }
\label{fig:xs_HA}
\end{centering}
\end{figure}
We show in Fig.~\ref{fig:xs_HA} the leading-order total cross section of the $e^+e^-\to ht\bar{t}$ process in the two limits, the pure CP-even ($h=H$) limit when $\xi_{htt}=0$, $\kappa_{htt}=\kappa_{hZZ}=1$, and the pure CP-odd ($h=A$) limit when $\xi_{htt}=\frac{\pi}{2}$, ~$\kappa_{htt}=1,~\kappa_{hZZ}$=0. $\sigma(Ht\bar{t})$ is about 0.28~fb at $\sqrt{s}=500$~GeV, reaching 2~fb at $\sqrt{s}\sim600$~GeV and stays above 2~fb until $\sqrt{s}\sim1$~TeV. {On the other hand, $\sigma(At\bar{t})$ is about 0.0045~fb (below the scale of Fig.~\ref{fig:xs_HA}) at $\sqrt{s}=500$~GeV,} rising quickly with energy, reaching 0.43~fb at $\sqrt{s}=1000$~GeV.

Because the CP-violating asymmetries appear as interference effects between CP-even and CP-odd amplitudes, {we show also the ratio of the two cross sections, $R=\sigma(At\bar{t})/\sigma(Ht\bar{t})$ in Fig.~\ref{fig:xs_HA}}. We can very roughly expect that the CP asymmetry is proportional to $\sqrt{R}$. The scale of $R$ is given along the right-hand vertical axis. It grows rapidly from 0.016 at $\sqrt{s}=500$~GeV to 0.047 at $\sqrt{s}=550$~GeV, reaching 0.2 at $\sqrt{s}=1000$~GeV.

\subsection{QCD corrections}

\begin{figure}[t]
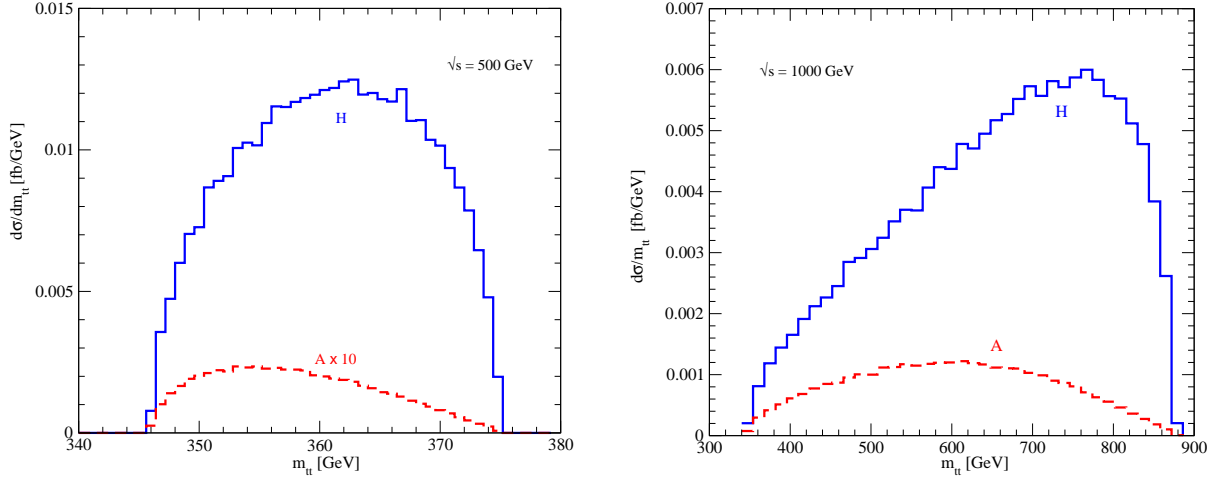

\vspace{0.5cm}
\begin{centering}
\begin{tabular}{c}
\includegraphics[width=0.42\textwidth]{mtt500.eps}
\hspace{0.5cm}
\includegraphics[width=0.43\textwidth]{mtt1000.eps}
\end{tabular}
\caption{$m_{t\bar{t}}$ distributions for $H$ and $A$ at $\sqrt{s}=500$~GeV (left) and 1000~GeV (right) in the leading order.}
\label{fig:mtt}
\end{centering}
\end{figure}

Shown in Fig.~\ref{fig:mtt} are the differential cross sections $d\sigma/ dm_{t\bar{t}}$ v.s.\ $m_{t\bar{t}}$ for $h=H$ (black solid curves) and for $h=A$ (red dashed curves) at $\sqrt{s}=500$~GeV (left) and at $\sqrt{s}=1000$~GeV (right) calculated in the leading order. The $h=A$ cross section at $\sqrt{s}=500$~GeV is multiplied by 10, in order to show the shape difference between the CP-even ($h=H$) and CP-odd ($h=A$) limits. It is clear from the two cases shown in the figure that the ratio of the CP-odd and CP-even amplitudes squared is large at low $m_{t\bar{t}}$, at all energies. This suggests that the sensitivity to CP asymmetry is high at low $m_{t\bar{t}}$, and hence the corrections including topponium formation can have significant impacts on our study of CP violation in the top Yukawa coupling. 

\begin{figure}[b]
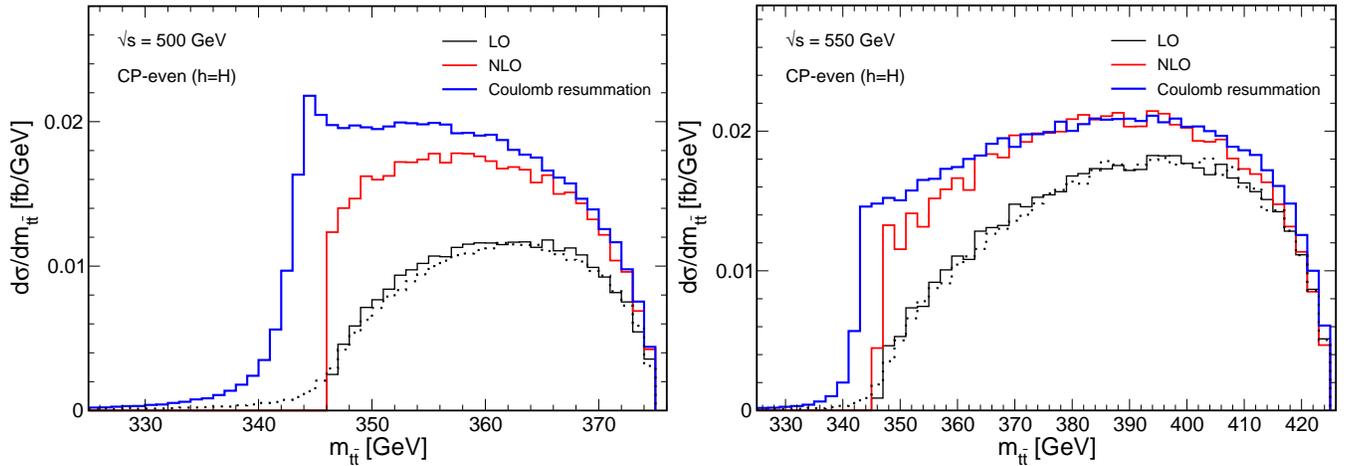

\vspace{0.8cm}
\includegraphics[width=0.49\textwidth]{dSigdMtt_S500.eps}
\includegraphics[width=0.49\textwidth]{dSigdMtt_S550.eps}
 \caption{
 $m_{t\bar{t}}$ distribution for the pure scalar case ($h=H$) at $\sqrt{s}=500$~GeV
 and 550~GeV.
 Black solid and red solid lines are the LO and NLO cross sections for
 the on-shell $t\bar{t}$, respectively,  
 and black dotted and blue solid lines are the Born-level and the
 Coulomb-resummed cross sections with top-quark off-shellness,
 respectively.
}\label{fig:mtt500550}
\end{figure}

We show in Fig.~\ref{fig:mtt500550} the differential cross section $d\sigma/d m_{t\bar{t}}$ for the SM Higgs boson ($h=H$) with QCD corrections.
NLO QCD corrections to the process are evaluated by using
MadGraph5$\_{\rm aMC@NLO}$~\cite{Alwall:2014hca}. 
In addition, we also consider the corrections by Coulomb
resummation~\cite{Hagiwara:2016rdv,Farrell:2005fk,Farrell:2006xe, Yonamine:2011jg}. 
According to
Refs.~\cite{Sumino:2010bv, Yonamine:2011jg}, we
estimate the Coulomb resummation corrections as follows. 
First, we evaluate the Born-level helicity amplitudes by including the
decays of top-quarks, $e^+ e^- \to ht\bar{t} \to hbW^+\bar bW^-$, in
which the top-quarks can be off-shell and $m_{t\bar{t}}$ can be below $2m_t$.
Then, the amplitudes are corrected by multiplying with the $S$-wave
Green function for non-relativistic $t\bar{t}$ with the energy $E =
m_{t\bar{t}} - 2m_t$. 
This prescription is justified by the fact that near the $t\bar{t}$
threshold where the Coulomb resummation is important the amplitudes are
dominated by the $S$-wave component.
To evaluate the Green function we employ the NLO QCD potential with
the renormalization scale $\mu=40$~GeV. 
Finally, the squared amplitudes are corrected by a hard correction
factor $K\simeq0.7$ which is numerically extracted by matching with the
NLO cross section at an intermediate $m_{t\bar{t}}$.

In Fig.~\ref{fig:mtt500550}, we show the differential cross section for the pure scalar
case $(h=H)$ at $\sqrt{s} = 500$~GeV (left) and 550~GeV (right). 
LO and NLO cross sections for the on-shell $t\bar{t}$ limit are plotted by
black solid and red solid lines, respectively. 
Born-level and the Coulomb-resummed cross sections with top-quark
off-shellness are plotted by black dotted and blue solid lines,
respectively. 
At $\sqrt{s}=500$~GeV, the NLO corrections enhance the total cross
section by around 60\%, while the Coulomb resummation (including the NLO effects) enhances the cross section by about 120\%.
At $\sqrt{s}=550$~GeV, the total cross section is enhanced by 20\% and
40\%, by the NLO and Coulomb resummation corrections, respectively. 

\begin{figure}[b]
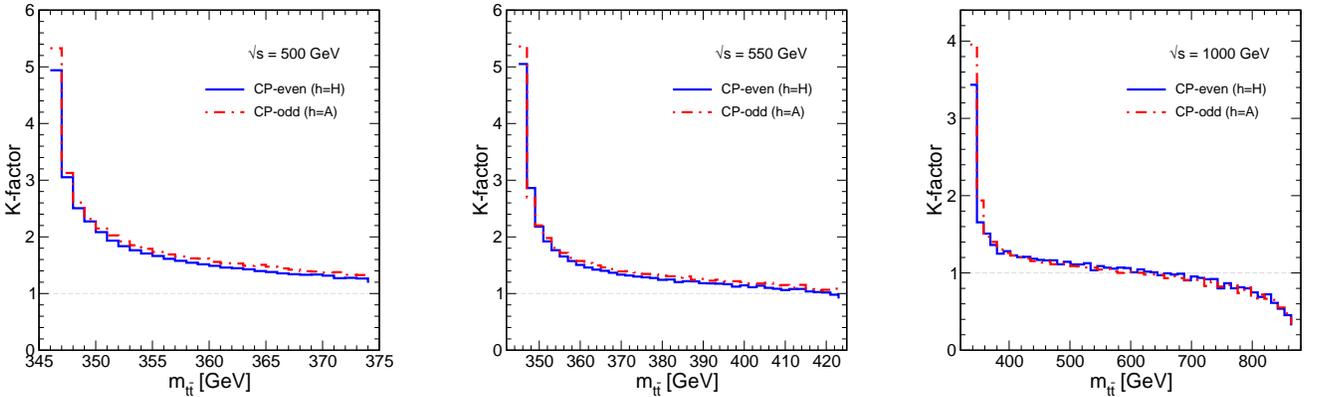

\vspace{0.8cm}
\begin{tabular}{c}
\includegraphics[width=0.29\textwidth]{Kfac500.eps}
\hspace{0.8cm}
\includegraphics[width=0.28\textwidth]{Kfac550.eps}
\hspace{0.8cm}
\includegraphics[width=0.28\textwidth]{Kfac1000.eps}
\end{tabular}
 \caption{
$m_{t\bar{t}}$ dependence of the NLO QCD $K$-factors in the $e^+e^-\to
 ht\bar{t}$ process for the CP-even ($h=H$) and CP-odd ($h=A$) limits at
 $\sqrt{s}=500$~GeV, 550~GeV, and 1000~GeV. 
 }\label{fig:kfac}
\end{figure}
\begin{table}[t]
 \begin{center}
  \begin{tabular}{|r|cc|cc|}
   \hline
   $\sqrt{s}$~~~& \multicolumn{2}{c|}{CP-even ($h=H$)} & \multicolumn{2}{c|}{CP-odd ($h=A$)} \\
   $[$GeV$]$ & NLO & Coulomb resum & NLO & Coulomb resum \\
   \hline \hline
   500 & 1.57 & 2.18 & 1.76 & 2.89 \\
   550 & 1.26 & 1.39 & 1.36 & 1.60 \\
   1000 & 0.95 & -    & 1.05 & -    \\
   \hline
  \end{tabular}
  \caption{$K$-factors by NLO and Coulomb-resummation corrections  to the
  total cross sections of the process $e^+e^-\to ht\bar{t}$ for the CP-even ($h=H$) and CP-odd ($h=A$) cases, at $\sqrt{s}$=500~GeV, 550~GeV, and 1000~GeV.} 
  \label{tab:nlo}
 \end{center}
\end{table}

In Fig.~\ref{fig:kfac}, we plot the ratio of the NLO cross section to the LO cross
section as a function of $m_{t\bar{t}}$ for the pure scalar ($h=H$) and pseudoscalar ($h=A$)
cases at $\sqrt{s} = 500$~GeV, 550~GeV, and 1000~GeV. 
We find the $K$-factors for the $m_{t\bar{t}}$ distribution are almost the
same for the scalar and pseudoscalar processes. 
QCD corrections are large near the threshold, $m_{t\bar{t}}\simeq 2m_t$,
because of the Coulomb singularity, but become almost flat and small at
large $m_{t\bar{t}}$.  

In Table~\ref{tab:nlo}, we list the $K$-factors of the total $ht\bar{t}$ production
cross section by QCD correction for the pure scalar ($h=H$) and
pseudoscalar ($h=A$) limits at $\sqrt{s}=500$~GeV, 550~GeV, and 1000~GeV.
An evaluation of the Coulomb resummation at $\sqrt{s}=1000$~GeV is
omitted because it has negligible effects. Both the NLO corrections and
Coulomb resummation (which contain NLO) corrections are largest at
$\sqrt{s}= 500$~GeV, still significant at 550~GeV, but small at
1000~GeV. The enhancement factors are larger for CP-odd case ($h=A$),
which is consistent with the softer $m_{t\bar{t}}$ distribution of the
$At\bar{t}$ process as shown in Fig.~\ref{fig:mtt}, where the $K$-factor
is large at small $m_{t\bar{t}}$ at all energies; see Fig.~\ref{fig:kfac}. The
NLO $K$-factor is smaller than unity for $h=H$ at $\sqrt{s}=1000$~GeV,
because the $m_{t\bar{t}}$ dependent $K$ factor becomes smaller than unity at
$m_{t\bar{t}}\geq650$~GeV above which the $m_{t\bar{t}}$ distribution is large;
see Fig.~\ref{fig:mtt}. 

\begin{table}[t]
 \begin{center}
  \begin{tabular}{|r|ccc|ccc|}
   \hline
 $\sqrt{s}$~~~  & \multicolumn{3}{c|}{$\sigma (Ht\bar{t})$ [fb]} & \multicolumn{3}{c|}{$\sigma(At\bar{t})$ [fb]} \\
   $[$GeV$]$ &LO& NLO & with Coulomb resum & LO&NLO & with Coulomb resum \\
   \hline \hline
   500 & 0.28&0.44 & 0.61 & 0.0045&0.0079& 0.0130 \\
   550 & 1.12&1.41& 1.56 &0.047 &0.064 & 0.075 \\
   1000 &2.08& 1.97 & -    & 0.43&0.45 & -   \\
   \hline
  \end{tabular}
  \caption{$e^+e^-\to ht\bar{t}$ cross sections at LO, NLO and with Coulomb-resummation corrections for the pure CP-even ($h=H$) and CP-odd ($h=A$) limits
  at  $\sqrt{s}=500$~GeV, 550~GeV, and 1000~GeV.} 
  \label{tab:cs_coulomb}
 \end{center}
\end{table}

Finally in Table~\ref{tab:cs_coulomb}, we show the total cross sections for the CP-even ($h=H$) and CP-odd ($h=A$) limits at LO, NLO with stable top quarks, and after Coulomb resummation including the off-shell top quark effects. Despite the factor of 3 enhancements (see Table~\ref{tab:nlo}), $\sigma(At\bar{t})$ remains tiny, 0.013~fb at 500~GeV. It grows by a factor of 6 to 0.075~fb at 550~GeV, which has a significant impact on our CP-violation search at $e^+e^-$ colliders.

{
Although the NLO and topponium corrections are quite large, the
NNLO effects to the total cross sections are expected to be marginal, 
since potentially large higher order corrections in the $t\bar{t}$ threshold regions are incorporated by Coulomb summation.
QCD corrections to the angular distributions or correlations of top quark decay
products, which may affect determination on $\xi_{htt}^{}$, are not studied in this paper. Detailed analysis is desired in the future. 
}

\section{Sensitivity of $e^+e^-\to ht\bar{t}$ experiments on $\xi_{htt}$}\label{sec:chisq}
In this section, we study the potential of $e^+e^-\to ht\bar{t}$ experiments to discover CP violation in the top Yukawa coupling at a future linear $e^+e^-$ collider by postulating a perfect detector with no {systematic} uncertainties. Because the measurement accuracy depends on the total number of produced events that determines the statistical errors, we first estimate the total cross section as a function of the parameters of our effective Lagrangian ($\kappa_{htt},\xi_{htt},\kappa_{hZZ}$) in subsection~\ref{subsec:cs_A}. In the next subsection~\ref{subsec:cs_B}, we explain in detail how we can calculate the full differential distribution including both semi-leptonic and hadronic decays of $t$ and $\bar{t}$, that are observable by a perfect detector but not as perfect as capable of distinguishing $\bar{d} ~(d)$ jets from $u~(\bar{u})$ jets. In subsection~\ref{subsec:cs_C}, we introduce a very simple estimator $\chi^2$ function that measures all differences in the observable distributions between experiment $(\kappa_{htt}^{\rm ex},\xi_{htt}^{\rm ex}$) and theory ($\kappa_{htt}^{\rm th},\xi_{htt}^{\rm th}$), and study the potential of rejecting $\xi_{htt}^{\rm th}=-\xi_{htt}^{\rm ex}$ (observation of CP violation) as a function of $|\xi_{htt}^{\rm ex}|.$

\subsection{Total cross section}\label{subsec:cs_A}
The total cross section of the process $e^+e^-\to ht\bar{t}$ depends not only on the parameters of ($\kappa_{htt},\xi_{htt},\kappa_{hZZ}$) but also on the center-of-mass energy $\sqrt{s}$ and the beam polarization. Since the NLO and topponium formation corrections are obtained only for the SM limit ($\kappa_{htt},\xi_{htt},\kappa_{hZZ}$)~=~(1,0,1) and for the purely CP-odd limit ($\kappa_{htt},\xi_{htt},\kappa_{hZZ})=(1,\pm\pi/2,0)$ for unpolarized beams at $\sqrt{s}=500$~GeV, 550~GeV, 1000~GeV, we make the following simple parameterization 
\begin{eqnarray}\label{eq:ABCD}
{\sigma_\alpha(\kappa_{htt},\xi_{htt},\kappa_{hZZ})}
&=&\sigma_H^{\rm NLO}[A_\alpha(\kappa_{htt}\cos\xi_{htt})^2+B_\alpha(\kappa_{htt}\cos\xi_{htt}\kappa_{hZZ})+C_{\alpha}(\kappa_{hZZ})^2]+\sigma_A^{\rm NLO}[D_\alpha(\kappa_{htt}\sin\xi_{htt})^2]\nonumber\\
&+&\sigma_H^{\rm topp.}[A^\prime_\alpha(\kappa_{htt}\cos\xi_{htt})^2+B^\prime_\alpha(\kappa_{htt}\cos\xi_{htt}\kappa_{hZZ})+C^\prime_{\alpha}(\kappa_{hZZ})^2]+\sigma_A^{\rm topp.}[D^\prime_\alpha(\kappa_{htt}\sin\xi_{htt})^2].\nonumber\\
\end{eqnarray}
Here $\sigma_{H}^{\rm NLO}$ and $\sigma_{A}^{\rm NLO}$ are obtained from Table~\ref{tab:nlo}, whereas {we quote the difference between} $\sigma^{\rm NLO}$ and the total cross section after taking account of Coulomb resummation  as the 'topponium' cross section,
\begin{eqnarray}
\sigma^{\rm topp.}=\sigma^{\rm with~ Coulomb ~sum}-\sigma^{\rm NLO}.
\end{eqnarray}
Those cross sections values are tabulated in Table~\ref{tab:cs}.
\begin{table}[t]
 \begin{center}   
  \begin{tabular}{|r|cccc|}
   \hline
   $[$GeV$]$ &$\sigma_{H}^{\rm NLO}$ &$\sigma_{H}^{\rm topp.}$ &$\sigma_{A}^{\rm NLO}$& $\sigma_{A}^{\rm topp.}$ \\
   \hline \hline
   500 & 0.44 & 0.17 &0.0079& 0.0051\\
   550 & 1.41& 0.15&0.064 &0.011 \\
   1000 &1.97& $-$ & 0.45    & $-$  \\
   \hline
  \end{tabular}
  \caption{$e^+e^-\to ht\bar{t}$ cross sections at NLO and for the topponium formation, whose sum gives the total cross section after Coulomb resummation in Table~\ref{tab:nlo}. $h=H$ for the SM Higgs, $h=A$ for the CP-odd Higgs with ($\kappa_{htt},\xi_{htt},\kappa_{hZZ})=(1,\pm\pi/2,0)$.}  
   \label{tab:cs}
 \end{center}
\end{table}
%
%%%%%
\begin{table}[b]
 \begin{center}
  \begin{tabular}{|rr|cccc||cccc|}
   \hline
   $\sqrt{s}$&&$A_\alpha$& $B_\alpha$& $C_\alpha$&$D_\alpha$&$A_\alpha^\prime$& $B_\alpha^\prime$& {$C_\alpha^\prime$}& {$D_\alpha^\prime$}\\
   \hline
 500 &$e_L$& 2.755&0.057 & 0.003 &2.779 &2.756 & 0.060&0.008&2.800 \\
 &$e_R$& 1.212&-0.029 & 0.002 & 1.221&1.208 & -0.036&0.004&1.200\\
  550 &$e_L$& 2.726&0.061& 0.009&2.766 &2.716 & 0.074 &0.029&2.773\\
  &$e_R$& 1.222&-0.025& 0.007&1.234 &1.202 & -0.043 &0.022&1.227\\
 1000 &$e_L$& 2.562&0.078& 0.084 &2.736 &$-$& $-$&$-$&$-$\\
  &$e_R$&1.199& 0.013 & 0.064   & 1.264&$-$ & $-$ &$-$&$-$  \\
   \hline  
  \end{tabular}
  \caption{$e^+e^-\to ht\bar{t}$ cross sections at LO, with NLO and Coulomb-resummation corrections for the pure scalar and pseudoscalar processes
  at  $\sqrt{s}=500$~GeV, 550~GeV, and 1000~GeV.} 
  \label{tab:ABCD}
 \end{center}
\end{table}
%%%%%
%%%%%
%%%%%
%
All the coefficients of our parameterization Eq.~(\ref{eq:ABCD}), which are tabulated in Table~\ref{tab:ABCD}, are obtained by using the LO matrix elements as follows. We calculate the total cross sections $\sigma_L^{}$ and $\sigma_R^{}$, respectively, for purely left-handed ($e_L$) and right-handed ($e_R$) beam in the LO for several sets of ($\kappa_{htt},\xi_{htt},\kappa_{hZZ}$) parameters, and obtain the parametrisation Eq.~(\ref{eq:ABCD}) with $\sigma_{H}^{\rm LO}$ and $\sigma_{A}^{\rm LO}$, $\sigma_{H}^{\rm topp.}=\sigma_{A}^{\rm topp.}=0$. We approximate the NLO corrections simply by replacing $\sigma_{H}^{\rm LO}$ and $\sigma_A^{\rm LO}$ by $\sigma_{H}^{\rm NLO}$ and $\sigma_{A}^{\rm NLO}$, respectively. We note here that this is not accurate, and the NLO corrections should be made separately for the $(\kappa_{htt}\cos\xi_{htt})^2$ and the $(\kappa_{hZZ})^2$ term as well as the interference term proportional to $(\kappa_{htt}\cos\xi_{htt}\kappa_{hZZ})$ in the future. For the topponium coefficients ($A_\alpha^\prime,B_{\alpha}^\prime,C_{\alpha^\prime},D_{\alpha}^\prime$), we calculate the LO cross sections at $m_{t\bar{t}}=2m_t+0.1$~GeV, just above the threshold, and fix all the coefficients, normalizing the total cross section to the topponium cross sections obtained in Section~\ref{sec:results} and tabulated in Table~\ref{tab:cs}. Accordingly, the coefficients in Table~\ref{tab:ABCD} are normalized as
\begin{eqnarray}
\sum_{\alpha=L,R}A_\alpha+B_\alpha+C_\alpha=\sum_{\alpha=L,R}D_\alpha=\sum_{\alpha=L,R}A^\prime_\alpha+B^\prime_\alpha+C^\prime_\alpha=\sum_{\alpha=L,R}D^\prime_\alpha=4
\end{eqnarray}
We believe that this is a good approximation to the $S$-wave topponium
formation, which dominates the coulomb resummation corrections.  

%
%%%%%
%%%%%
%%%%%
%%%%%
%%%%%

The cross section for partially polarized beams $(|P_e|<1~{\rm for}~ e^-,~|P_{\bar{e}}|<1~{\rm for}~e^+)$ is obtained from Eq.~(\ref{eq:ABCD}) as 
\begin{eqnarray}
\sigma(P_e,P_{\bar{e}})=\frac{(1-P_e)(1+P_{\bar{e}})}{4}\sigma_L+\frac{(1+P_e)(1-P_{\bar{e}})}{4}\sigma_R
\end{eqnarray}
and hence the unpolarized cross sections are 
\begin{eqnarray}
\sigma(0,0)=\frac{1}{4}(\sigma_L+\sigma_R).
\end{eqnarray}
Here, we give a few comments on the impacts of the beam polarization on the $ht\bar{t}$ coupling measurements. We first note from Table~\ref{tab:ABCD} that the ratio $D_\alpha/A_{\alpha}$ and $D_{\alpha}^\prime/A_{\alpha}^\prime$ are almost the same for $e_L$ and $e_R$, and hence the beam polarization has little impacts (besides the total number of events which can be increased by choosing $P_e<0<P_{\bar{e}}$) in the measurements of the sign of $\xi_{htt}$ that arise from the interference between the amplitudes {whose squares} give $A_\alpha$ and $D_\alpha$, respectively. 
The most significant effect of the beam polarization is in the value of
$B_\alpha$, the coefficients of the interference between two CP-even
amplitudes, those with the $ht\bar{t}$ coupling proportional to
$\kappa_{htt}\cos\xi_{htt}$ and those with the $hZZ$ coupling. Their
signs are opposite at $\sqrt{s}=500$~GeV and 550~GeV, while the magnitude is
different by a factor of six at $\sqrt{s}=1000$~GeV. Accordingly, we
find significant improvement in resolving
$\xi_{htt}=0~(\cos\xi_{htt}=1)$ and $\xi_{htt}=\pm\pi
~(\cos\xi_{htt}=-1)$ by using the beam polarization. However, because the beam polarization has little impacts on the CP-violation measurements, the following studies are performed for unpolarized beams ($P_e=P_{\bar{e}}=0$). We would like to study the impacts of the $e^+e^-$ beam polarizations in a separate report, in which the sensitivity should be compared with those from single top and $h$ production at the LHC~\cite{Demartin:2015uha}.

Throughout this study, we adopt 
\begin{eqnarray}
L=1000~{\rm fb}^{-1}
\end{eqnarray}
as a nominal integrated luminosity at each colliding $e^+e^-$ energy $\sqrt{s}=500$~GeV, 550~GeV, and 1000~GeV so that we can compare the impacts of increasing the beam energy. For unpolarized beams, we expect the following number of $ht\bar{t}$ events with unpolarized beams for the SM $(h=H)$:
\begin{eqnarray}
610~{\rm events~}&&~{\rm at}\quad\sqrt{s}= 500~{\rm GeV}\nonumber\\
1530~{\rm events~}&&~{\rm at}\quad\sqrt{s}=550~{\rm GeV}\nonumber\\
2000~{\rm events~}&&~{\rm at}\quad\sqrt{s}=1000~{\rm GeV}\,
\end{eqnarray}
\begin{figure}[t]
%\vspace{0.8cm}
\begin{centering}
\begin{tabular}{c}
\includegraphics[width=0.45\textwidth]{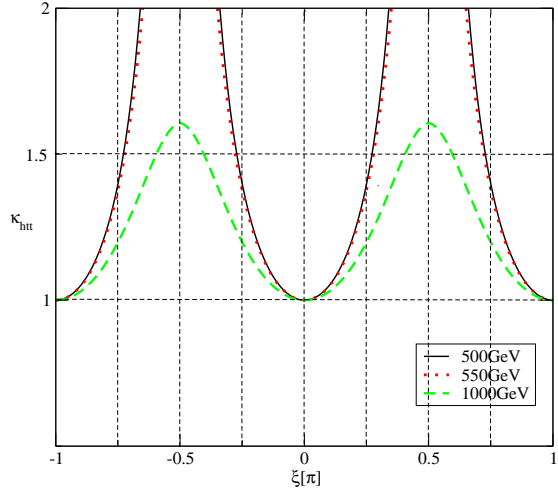}
\end{tabular}
\caption {The $\kappa_{htt}^{}$ value which reproduces the SM ($h=H$) cross section for $\xi_{htt}\neq0$, at $\sqrt{s}=500$~GeV (black solid), 550~GeV (red dotted) and 1000~GeV (green dashed). The results are for $\kappa_{hZZ}^{}=1$.}
\label{fig:k-xi}
\end{centering}
\end{figure}
\hspace{-0.18cm}which measures the total cross section with statistical errors of $4.0\%$, $2.6\%$, $2.2\%$, respectively. If we assume that these number of events are observed, {then} at each energy, the value of $(\kappa_{htt},\xi_{htt})$ are constrained to lie on the curves shown in {Fig.~\ref{fig:k-xi}}. We set $\kappa_{hZZ}=1$ throughout our studies. The statistical error on $\kappa_{htt}$ is about half the cross section error, $\sim2.0\%$,$\sim1.3\%$, $\sim1.1\%$, respectively, for $\sqrt{s}=500$~GeV, 550~GeV, 1000~GeV, which are too small to show in the plot. It is clear that the measurement of the total cross section at two different energies can constrain a region of $\kappa_{htt}$ and $\xi_{htt}$, and more importantly, the LHC experiments on $ht\bar{t}$ and $ht$, $h\bar{t}$ productions will also give such constraints. We therefore assume that the SM Higgs cross section is observed at a given $e^+e^-$ collision energy, and study the capability of distribution studies to resolve the sign of $\xi_{htt}$ as a function of $|\xi_{htt}|$. Therefore, one should be careful when comparing $\xi_{htt}=\frac{\pi}{4}$ against $\xi_{htt}=-\frac{\pi}{4}$ at $\sqrt{s}=500$~GeV, since it is not $\kappa_{htt}=1$ but $\kappa_{htt}=1.4$ which gives the same number of events as in the SM.

\subsection{Differential cross sections}\label{subsec:cs_B}
The differential cross sections including $t$ and $\bar{t}$ decay distributions are obtained as in Eqs.~(\ref{eq:diff_xs_set}) by using the density matrix formalism. In this subsection we explain how we treat the hadronic decays of $t$ and $\bar{t}$, QCD corrections to the differential cross sections which strongly depend on $m_{t\bar{t}}$, possible beam polarizations, and how we take account of the probability that $t$ and $\bar{t}$ cannot be distinguished uniquely when both of them decay hadronically.

First, we introduce the following generic form of the $t$ and $\bar{t}$ decay density matrices
\begin{subequations}\label{eq:dens_matr}
\begin{align}
{\rho}&=\frac{6(1-\bar{x})}{\left(1-\frac{m_W^2}{m_t^2}\right)\left(1+2\frac{m_W^2}{m_t^2}\right)}
\left(\begin{array}{cc}
\bar{x}+\bar{x}_z&\bar{x}_x+i\bar{x}_y\\
\bar{x}_x-i\bar{x}_y&\bar{x}-\bar{x}_z
\end{array}\right),\\
  {\bar{\rho}}&=\frac{6(1-x)}{\left(1-\frac{m_W^2}{m_t^2}\right)\left(1+2\frac{m_W^2}{m_t^2}\right)}
  \left(\begin{array}{cc}
{x+x_z}&{x_x-ix_y}\\
{x_x+ix_y}&{x-x_z}
\end{array}\right),
\end{align}
\end{subequations}
respectively, where
\begin{subequations}
\begin{align}
\bar{x}^\mu=\frac{2p_{\bar{\ell}}^\mu}{m_t}=(\bar{x},\bar{x}_x,\bar{x}_y,\bar{x}_z),\\
{x}^\mu=\frac{2p_{{\ell}}^\mu}{m_t}=({x},{x}_x,{x}_y,{x}_z),\noindent\,
\end{align}\label{eq:xmu}
\end{subequations}
\hspace{-0.18cm}are normalized 4-momenta of $\bar{\ell}~(\bar{d}$ or $\bar{s}$) and
${\ell}~ (d$ or $s$), respectively, in the $t$ and $\bar{t}$ rest
frame. They are expressed in terms of the $b~(\bar{b})$ angles
$\theta_b^\ast,\phi_b^\ast~(\theta_{\bar{b}},\bar{\phi}_{\bar{b}}^\ast)$
in the $t~(\bar{t})$ rest frame, the $\bar{\ell},\bar{d},\bar{s}$ angles
$\bar{\theta}^{\ast\ast},\bar{\phi}^{\ast\ast}$ in the $W^+$ rest
frame, and the $\ell,d,s$ angles $\theta^{\ast\ast},\phi^{\ast\ast}$
in the $W^-$ rest frame. Details are given in Appendix ~\ref{sec:appA}. The density
matrices in Eq.~(\ref{eq:dens_matr}) are normalized as 
\begin{subequations}
\begin{align}
\int\rho_{\sigma\sigma^\prime}\frac{d\cos\theta_b^\ast d\phi_b^\ast}{4\pi}\frac{d\cos\bar{\theta}^{\ast\ast} d\bar{\phi}^{\ast\ast}}{4\pi}=\delta_{\sigma\sigma^\prime}\\
\int\bar{\rho}_{\bar{\sigma}\bar{\sigma}^\prime}\frac{d\cos\theta_{\bar{b}}^\ast d\phi_{\bar{b}}^\ast}{4\pi}\frac{d\cos{\theta}^{\ast\ast} d{\phi}^{\ast\ast}}{4\pi}=\delta_{\bar{\sigma}\bar{\sigma}^\prime}
\end{align}
\end{subequations}
We use the density matrices Eq.~(\ref{eq:dens_matr}) for all leptonic decays. Indeed, integration of Eq.~(\ref{eq:dens_matr}) over the $b~({\bar{b}})$ angles for fixed angles of $\bar{\ell}~(\ell)$ in the $t~(\bar{t})$ rest frame reproduce the well known formula Eq.~(\ref{eq:drho2}) for $B_{\ell}=1$.

 For hadronic decays, we introduce the probability $P_{\bar{d}u}$ with which we identify $\bar{d}$ from $u$ (${d}$ from $\bar{u}$), or $\bar{s}$ from $c$ ($s$ from $\bar{c}$) in $W^-(W^+)$ decays
\begin{subequations}\label{eq:dens_matr_h}
\begin{align}
\rho^h&=\frac{1+P_{\bar{d}u}}{2}\rho(\theta_b^\ast,\phi_b^\ast,\bar{\theta}^{\ast\ast},\bar{\phi}^{\ast\ast})+\frac{1-P_{\bar{d}u}}{2}\rho(\theta_b^\ast,\phi_b^\ast,\pi-\bar{\theta}^{\ast\ast},\pi+\bar{\phi}^{\ast\ast})\\
\bar{\rho}^h&=\frac{1+P_{\bar{d}u}}{2}\bar{\rho}(\theta_{\bar{b}}^\ast,\phi_{\bar{b}}^\ast,{\theta}^{\ast\ast},{\phi}^{\ast\ast})+\frac{1-P_{\bar{d}u}}{2}\rho(\theta_{\bar{b}}^\ast,\phi_{\bar{b}}^\ast, \pi-{\theta}^{\ast\ast},\pi+{\phi}^{\ast\ast})
\end{align}
\end{subequations}
 where we assume that $b$ and $\bar{b}$ momenta are uniquely identified. 
With the probability $(1+P_{\bar{d}{u}})/2$, we identify $\bar{d}$ or $\bar{s}$ ($d$ or $s$) correctly in $W^+(W^-)$ decays, whereas with the probability $(1-P_{\bar{d}u})/2$, we misidentify $u$ or $s$ ($\bar{u}$ or $\bar{s}$) for $\bar{d}$ or $\bar{s}$ ($d$ or $s$). 
 In our analysis, we set
 \begin{eqnarray}
 P_{\bar{d}u}=P_{\bar{s}c}=0
 \label{eq:P0}
 \end{eqnarray}
 for simplicity, but $P_{\bar{s}c}$ may be significant. The decay
 density matrices $\rho^h$ and $\bar{\rho}^h$ in Eqs.~(\ref{eq:dens_matr_h}) keep the maximum information given by the matrix elements. If we ignore the top spin sensitivity in the $\bar{d}$ or $\bar{s}$ ($d$ or $s$) distributions by integrating over the $W$ decay angles $(\bar{\theta}^{\ast\ast},\bar{\phi}^{\ast\ast})$, only poor resolution power of $b$ angular variables $\theta_b^\ast,\phi_b^\ast$ ($\theta_{\bar{b}}^\ast,\phi_{\bar{b}}^\ast$) remains.
 
 By using $\rho$ and $\bar{\rho}$ in Eq.~(\ref{eq:dens_matr}) for leptonic decays and $\rho^h$ and $\bar{\rho}^h$ in Eq.~(\ref{eq:dens_matr_h}) for hadronic decays, the differential cross sections are expressed as follows
 \begin{eqnarray}\label{eq:dsigdphi}
 \frac{d\sigma}{d\Phi}=\sum_{\sigma}\sum_{\bar{\sigma}}\sum_{\sigma^\prime}\sum_{\bar{\sigma}^\prime}\left\{{\cal S}_{\sigma\bar{\sigma}\sigma^\prime\bar{\sigma}^\prime}\left[B_\ell^2\rho_{\sigma\sigma^\prime}\bar{\rho}_{\bar{\sigma}\bar{\sigma}^\prime}+B_\ell B_h(\rho_{\sigma\sigma^\prime}\bar{\rho}_{\bar{\sigma}\bar{\sigma}^\prime}^h+\rho^h_{\sigma\sigma^\prime}\bar{\rho}_{\bar{\sigma}\bar{\sigma}^\prime})\right]+B_h^2{\cal S}^{hh}_{\sigma\bar{\sigma}\sigma^\prime\bar{\sigma}^\prime}\rho^{h^\prime}_{\sigma\sigma^\prime}\bar{\rho}_{\bar{\sigma}\bar{\sigma}^\prime}^{h^\prime}\right\}
 \end{eqnarray}
 with the reduced phase space
 \begin{eqnarray}
 d\Phi=d\Phi_{htt}\frac{d\cos\theta_b^\ast d\phi_b^\ast}{4\pi}\frac{d\cos\bar{\theta}^{\ast\ast}d\bar{\phi}^{\ast\ast}}{4\pi}\frac{d\cos\theta_{\bar{b}}^\ast d\phi_{\bar{b}}^\ast}{4\pi}\frac{d\cos{\theta}^{\ast\ast}d{\phi}^{\ast\ast}}{4\pi}.
 \end{eqnarray}
 In this report, we do not distinguish lepton or quark flavours, and adopt
\begin{subequations}
\begin{align}
B_\ell&=\sum_{\ell=e,\mu,\tau}B(t\to b\bar{\ell}\nu_{\ell})\simeq0.33,\\
B_h&=B(t\to b\bar{d}u)+B(t\to b\bar{s}c)=1-B_\ell\simeq0.67.
\end{align}
\end{subequations}
 Production density matrices are expressed in terms of the helicity amplitudes as 
 \begin{eqnarray}
 {\cal S}_{\sigma\bar{\sigma}\sigma^\prime\bar{\sigma}^\prime}=\frac{1}{2s}\sum_\alpha\frac{1+\alpha P_e}{2}\frac{1-\alpha P_{\bar{e}}}{2}M_{\alpha\sigma\bar{\sigma}}M_{\alpha\sigma^\prime\bar{\sigma}^\prime}^\ast
 \end{eqnarray}
 where $-1<P_e,P_{\bar{e}}<1$ denote $e$ and $\bar{e}$ longitudinal beam polarizations. 
 
There is one subtlety when both $t$ and $\bar{t}$ decay hadronically. In this case, identification of $t$ and $\bar{t}$ can be ambiguous, and we introduce the probability $P_{t\bar{t}}$ with which $t$ and $\bar{t}$ can be identified correctly. In this report, we adopt
\begin{eqnarray}\label{eq:ptt0.4}
P_{t\bar{t}}=0.4
\end{eqnarray}
 which is approximately twice the semi-leptonic decay branching fraction of $B$ mesons. Charge discrimination of hadronic jets from $W^+$ and $W^-$ decays may also help. With this probability, the distribution proportional to $B_h^2\simeq0.45$ in Eq.~(\ref{eq:dsigdphi}) should be 
 \begin{eqnarray}
 {{\cal S}^{hh}_{\sigma\bar{\sigma}\sigma^\prime\bar{\sigma}^\prime}
 \rho_{\sigma\sigma^\prime}^{h^\prime}\bar{\rho}_{\bar{\sigma}\bar{\sigma}^\prime}^{h^\prime}}
 &=&
 \frac{1+P_{t\bar{t}}}{2}{\cal S}_{\sigma\bar{\sigma}\sigma^\prime\bar{\sigma}^\prime}
 (\hat{\theta},\hat{\phi})
 \rho^h_{\sigma\sigma^\prime}
  (\theta_b^\ast,\phi_b^\ast,\bar{\theta}^{\ast\ast},\bar{\phi}^{\ast\ast})
 \bar{\rho}^h_{\bar{\sigma}\bar{\sigma}^\prime}
 (\theta_{\bar{b}}^\ast,\phi_{\bar{b}}^\ast,{\theta}^{\ast\ast},{\phi}^{\ast\ast})
 \nonumber\\
 &+&
\frac{1-P_{t\bar{t}}}{2}{\cal S}_{\sigma\bar{\sigma}\sigma^\prime\bar{\sigma}^\prime}
(\pi-\hat{\theta},\pi+\hat{\phi})
\rho_{\sigma\sigma^\prime}^h
(\theta_{\bar{b}}^\ast,\phi_{\bar{b}}^\ast,{\theta}^{\ast\ast},{\phi}^{\ast\ast})
\bar{\rho}_{\bar{\sigma}\bar{\sigma}^\prime}^h
(\theta_b^\ast,\phi_b^\ast,\bar{\theta}^{\ast\ast},\bar{\phi}^{\ast\ast})
 \end{eqnarray}
 In the last term with the probability $(1-P_{t\bar{t}})/2$, $t$ and $\bar{t}$ are misidentified, not only the $t$ angles $(\hat{\theta},\hat{\phi})$ in the $t\bar{t}$ rest frame are replaced by those of $\bar{t}$ angles $(\pi-\hat{\theta},\pi+\hat{\phi})$, but also all the 4-angles of $t$ and $\bar{t}$ decays are exchanged. Although this looks complicated, it is straightforward to implement it in a numerical program, and we can keep the maximum surviving information of the matrix elements.
 
 Finally, we implement the perturbative NLO corrections and the
 topponium contribution as follows. For the NLO corrections, we find in
 Section~\ref{sec:results} that the $m_{t\bar{t}}$-dependent $K$-factors are almost
 identical to the CP-even and CP-odd cases, see Fig.~\ref{fig:kfac}. We
 therefore ignore their small differences, and multiply the $m_{t\bar
 t}$-dependent $K$-factor of the CP-even ($h=H$) cross section to our
 differential cross section in Eq.~(\ref{eq:dsigdphi}) 
 \begin{eqnarray}\label{eq:diff_kfac}
 \frac{d\sigma}{d\Phi}\to K(m_{t\bar{t}})\frac{d\sigma}{d\Phi}.
 \end{eqnarray}
 We confirm that this simple prescription reproduces all the NLO cross sections listed in Table~\ref{tab:cs}.
 
 For the topponium contribution at $\sqrt{s}=500$~GeV and 550~GeV, we
 evaluate the differential cross section Eq.~(\ref{eq:dsigdphi}) at
 $m_{t\bar{t}}=2m_t+0.1$~GeV, and normalize the total integral such that
 it agrees with the production cross section of topponium + Higgs listed in Table~\ref{tab:cs_coulomb}, by multiplying the factor
 \begin{eqnarray}
 \frac{\sigma_{H}^{\rm topp.}}{\sigma(m_{t\bar{t}}=2m_t+0.1~{\rm GeV})}.
 \end{eqnarray}
 With this prescription, we obtain all the correlated decays of both $t$
 and $\bar{t}$ decays, reproducing the results of
 Ref.~\cite{Hagiwara:2016rdv}.
 
Sophisticated simulation program with topponium formation and decays is needed to evaluate the cross sections and distributions for the SM and its extensions. We believe, however, that the model dependences are approximated reasonably well by using the leading-order matrix elements as outlined in this subsection.
\subsection{Results}\label{subsec:cs_C}
As an estimator for the sensitivity of our CP-violation measurement in $e^+e^-\to ht\bar{t}$ process, we introduce the following $\chi^2$ function
\begin{eqnarray}
\chi^2(\kappa_{htt},\xi_{htt})=L\int d\Phi\left(\frac{d\sigma_{ex}/d\Phi-d\sigma_{th}(\kappa_{htt},\xi_{htt})/d\Phi}{\sqrt{d\sigma_{th}(\kappa_{htt},\xi_{htt})/d\Phi}}\right)^2,
\label{eq:chisq-3bd}
\end{eqnarray}
Here $d\sigma_{ex}/d\Phi$ represents the observed experimental cross section calculated by assuming that a set of parameters  ($\kappa_{htt}^{\rm ex}$, $\xi_{htt}^{\rm ex}$) are true and $d\sigma_{th}/d\Phi$ is calculated for an arbitrary set of
($\kappa_{htt}^{}$, $\xi_{htt}^{}$) values at each energy.  $L$ is the integrated luminosity of the process for unpolarized beams. Throughout our study, we set 
\begin{eqnarray}
L=1000~{\rm fb}^{-1},~~P_e=P_{\bar{e}}=0,
 \end{eqnarray}
at all energies. The $\chi^2$ function in Eq.~(\ref{eq:chisq-3bd}) accounts for all possible difference between data $d\sigma_{ex}/d\Phi$ and theory $d\sigma_{th}/d\Phi$ at all kinematical configuration $\Phi$, with the corresponding statistical error proportional to $(Ld\sigma_{\rm th}/d\Phi)^{-1/2}$. It maybe regarded as an ultimate possible sensitivity for a perfect detector with infinite resolution and no errors. Because we calculate the data $d\sigma/d\Phi$ by using our theoretical formula without fluctuations, it is obvious that 
\begin{eqnarray}
\chi^2_{\rm min}=\chi^2(\kappa_{htt}^{ex},\xi_{htt}^{ex})=0
\end{eqnarray}
When we fix the test value of $\xi_{htt}$, the minimum becomes
\begin{eqnarray}
\chi^2_{\rm min}(\xi_{htt})=\chi^2\left(\kappa_{htt}(\xi_{htt}),\xi_{htt}\right)
\end{eqnarray}
where $\kappa_{htt}(\xi_{htt})$ is the value which gives the same total cross section with the SM. The trajectory of $\kappa_{htt}(\xi_{htt})$ has been given in Fig.~\ref{fig:k-xi} for $\sqrt{s}=500$~GeV, 550~GeV, and 1000~GeV.
\begin{figure}[b]
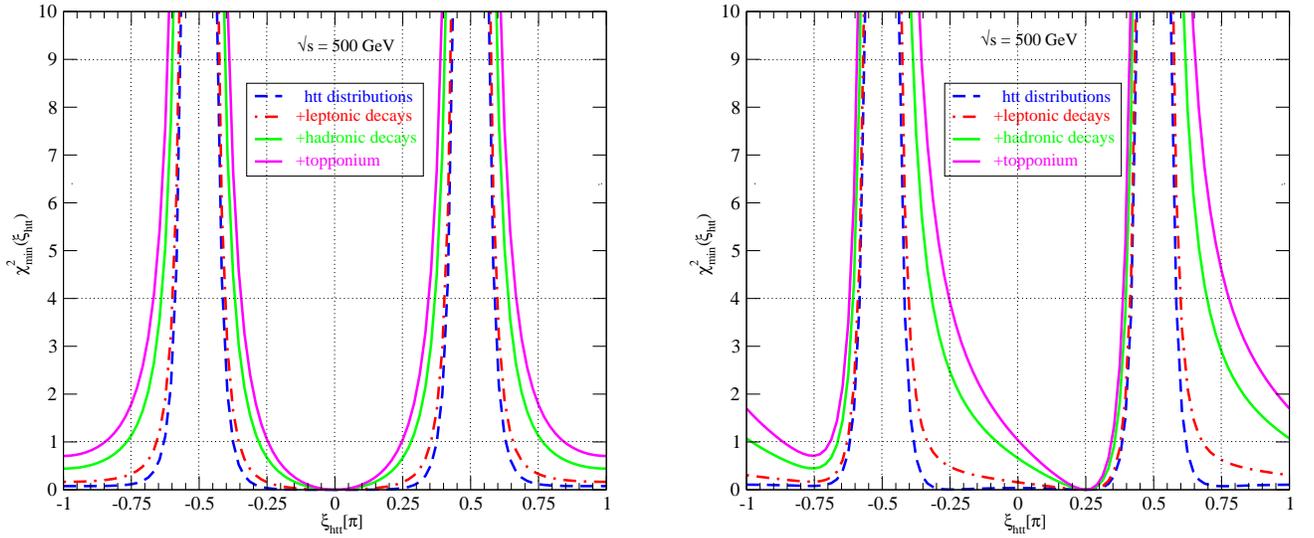

\vspace{0.8cm}
\begin{centering}
\begin{tabular}{c}
\includegraphics[width=0.45\textwidth]{chisq500_xi0.eps}
\hspace{0.8cm}
\includegraphics[width=0.45\textwidth]{chisq500_xi025.eps}
\end{tabular}
\caption {The $\chi^2$ plot  at $\sqrt{s}=500$~GeV. $\xi_{htt}^{\rm ex}=0$ for the left plot, $\xi_{htt}^{ex}=\pi/4$ for the right plot.}
\label{fig:chisq500}
\end{centering}
\end{figure}

In the left panel of Fig.~\ref{fig:chisq500}, we show $\chi^2_{\rm min}(\xi_{htt})$ as a function of $\xi_{htt}$ when the SM distribution $(\kappa_{htt},\xi_{htt})=(1,0)$ is assumed for $d\sigma_{ex}/d\Phi$. The blue dashed curve is obtained when only the $ht\bar{t}$ distributions are measured. In our differential cross section formula in Eq.~(\ref{eq:diff_kfac}) with $K(m_{t\bar{t}})$ for NLO correction, these limits are obtained simply by replacing all our decay density matrices by a unit matrix:
\begin{eqnarray}
\rho_{\sigma\sigma^\prime}=\rho_{\sigma\sigma^\prime}^h=\delta_{\sigma\sigma^\prime},
~~~~
 \bar{\rho}_{\bar{\sigma}\bar{\sigma}^\prime}=\bar{\rho}_{\bar{\sigma}\bar{\sigma}^\prime}^h=\delta_{\bar{\sigma}\bar{\sigma}^\prime}.
\end{eqnarray}
As compared to the simple calculation of $e^+e^-\to ht\bar{t}$ differential cross section, we find slightly smaller $\chi^2_{\rm min}$ because of finite $t\bar{t}$ discrimination probability $P_{t\bar{t}}=0.4$ when both $t$ and $\bar{t}$ decay hadronically.
The red dash-dotted curve shows our results when leptonic decay angular correlations are assumed. This limit is obtained in our $d\sigma/d\Phi$ formula by setting only the hadronic decay density matrix to be a unit matrix.
\begin{eqnarray}
\rho_{\sigma\sigma^\prime}^h=\delta_{\sigma\sigma^\prime},~\bar{\rho}_{\bar{\sigma}\bar{\sigma}^\prime}^h=\delta_{\bar{\sigma}\bar{\sigma}^\prime}.
\end{eqnarray}
Only small improvements are found over the $ht\bar{t}$ distributions only case. 
This is mainly because of the smallness of the branching fraction factor of the dilepton case $B_\ell^2\sim0.11$ even including the $\ell=\tau$ and $\ell=\bar{\tau}$ modes, even though they give full $t$ and $\bar{t}$ decay angular correlations. It is also because the single ($t$ or $\bar{t}$) decay angular distributions with the probability of $2B_\ell (1-B_\ell)\sim0.44$ have relatively small sensitivity to the CP phase, $\xi_{htt}$.
The green solid curve is obtained by including both leptonic and hadronic decay angular correlations by using our full differential cross sections $d\sigma/d\Phi$ in Eq.~(\ref{eq:diff_kfac}). 
The significant improvement over the red dash-dotted curve shows that the angular correlation between semi-leptonic and hadronic decays keep the resolving power even with the assumption of $P_{\bar{d}u}=P_{\bar{s}c}=0$ in Eq.~(\ref{eq:P0}).
Finally, the red solid curve is obtained by adding the topponium contributions. 
Although the full decay angular distribution studies based on our formalism improve the measurement significantly, it is clear from the plot that the SM data at $\sqrt{s}=500$~GeV is consistent with $|\xi_{htt}|=\pi/4$ {at 1$\sigma$}, even with 1000~fb$^{-1}$. 

In the right panel of Fig.~\ref{fig:chisq500}, we show the results when the data follow the prediction of the maximum CP phase at $(\kappa_{htt}^{ex},\xi_{htt}^{ex})=(1.4,\pi/4)$, where the total cross section is the same as the SM. 
Here the {$\chi_{\rm min}^2$} value at $\xi_{htt}=-\xi_{htt}^{\rm ex}=-\pi/4$ gives the sensitivity to CP violation. Apparently, the $ht\bar{t}$ distribution has no sensitivity (see the blue dashed curve), as may be expected from Fig.~\ref{fig:azm} in Section II, where the difference between $\xi_{htt}=\pi/4$ and $-\pi/4$ is tiny for the sum of squared of all the amplitudes. The inclusion of leptonic decays improves (red dash-dotted line), and the impact of including hadronic decays (green solid line) is quite significant. Adding the topponium contributions, the $\chi^2$ value reaches 4, or we might find a 2$\sigma$ hint of CP violation.

We find the results discouraging, because this result is for the largest possible CP phase, with 1000~fb$^{-1}$, and a prefect detector with no fluctuations in data are assumed in the analysis. This leads us to re-examine the target energy of a linear collider. As explained in Section~\ref{sec:helicity} , the disappointing results at $\sqrt{s}=500$~GeV are probably unavoidable, because the small ratio $R=\sigma(At\bar{t})/\sigma(Ht\bar{t})={0.016}$ at $\sqrt{s}=500$~GeV shown in Fig.~\ref{fig:xs_HA} implies that the CP-odd amplitudes are much smaller than the CP-even amplitudes. Our analytic form of the helicity amplitudes obtained in Section II  in Eq.~(\ref{eq:amp_ass}) and Eqs.~(\ref{eq:lampmmp}-\ref{eq:0pmpm}) confirms that all the amplitudes proportional to $\sin\xi_{htt}$ are either proportional to $\hat{\beta}$, the $t$ velocity in the $t\bar{t}$ rest frame, $\beta$, the velocity of the $t\bar{t}$ system in the $e\bar{e}$ rest frame, or $D_t^1-D_t^2$, the difference between the two top quark propagator factors, which are all strongly suppressed at energies near $ht\bar{t}$ threshold, $m_h+2m_t~=~471$~GeV.

\begin{figure}[b]
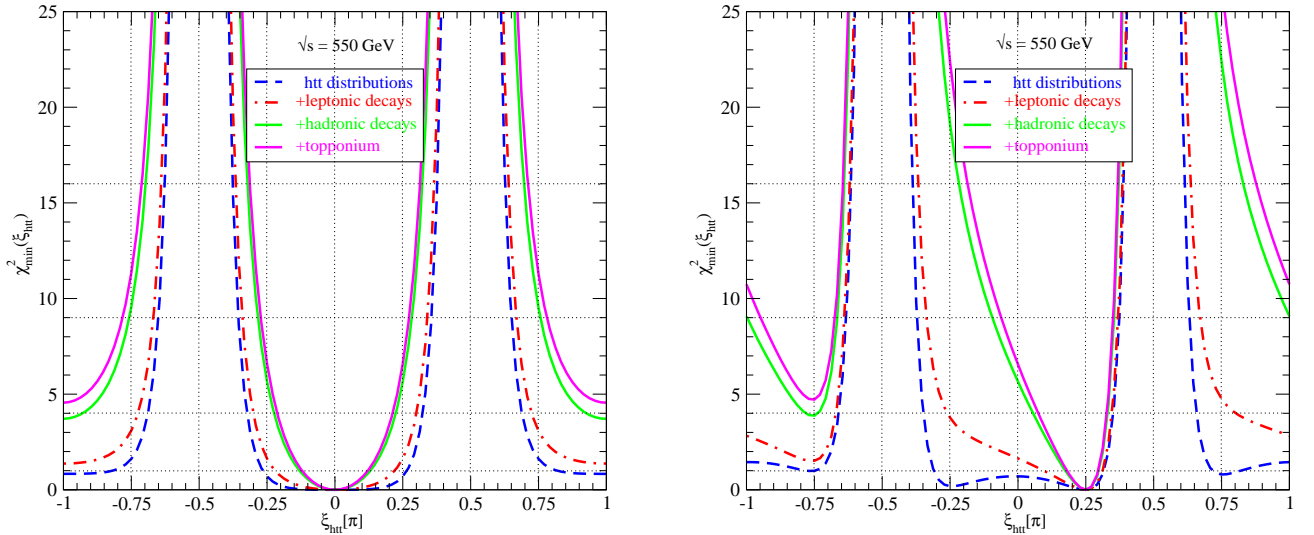

\vspace{0.8cm}
\begin{centering}
\begin{tabular}{c}
\includegraphics[width=0.45\textwidth]{chisq550_xi0.eps}
\hspace{0.8cm}
\includegraphics[width=0.45\textwidth]{chisq550_xi025.eps}\\
\end{tabular}
\caption {The $\chi^2$ plot at $\sqrt{s}=550$~GeV assuming $\xi_{htt}^{ex}=0$ as the observed data (left panel) and $\xi_{htt}^{ex}=\pi/4$ as the observed data (right panel).}
\label{fig:chisq550}
\end{centering}
\end{figure}
\begin{figure}[b]
\vspace{0.8cm}
\begin{centering}
\begin{tabular}{c}
\includegraphics[width=0.45\textwidth]{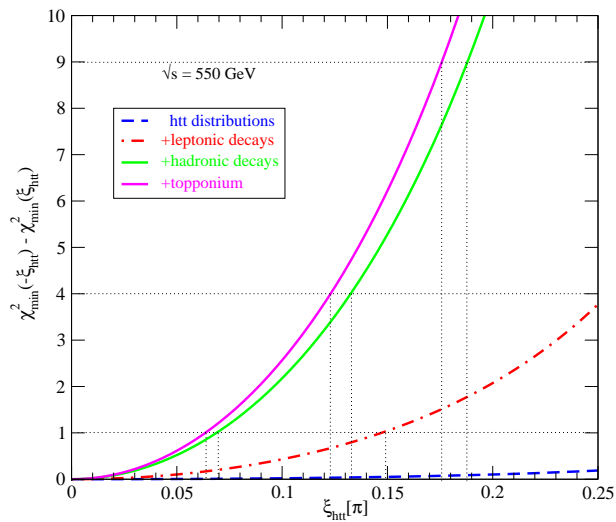}
\end{tabular}
\caption { Sensitivity to CP violation is shown by $\chi_{\rm min}^2 (\xi_{htt}=-\xi_{htt}^{\rm ex})-\chi^2_{\rm min}(\xi_{htt}=\xi_{htt}^{\rm ex})$ plotted against $\xi_{htt}^{\rm ex}$, the assumed true value of the CP phase at $\sqrt{s}=550$~GeV.}
\label{fig:chisq550CP}
\end{centering}
\end{figure}
We therefore examine the possibility of upgrading the target energy by
$10\%$ to 550~GeV, where the ratio
$R=\sigma(At\bar{t})/\sigma(Ht\bar{t})$ increases to 0.047 $\sim
(0.22)^2$. The results are shown in Fig.~\ref{fig:chisq550}. In the left
panel, we show $\chi_{\rm}^2(\xi_{htt})$ when the SM $(\kappa_{htt}^{\rm
ex},\xi_{htt}^{\rm ex})=(1,0)$ is assumed for the data. Again, the
$ht\bar{t}$ distributions (modified for the probability
$B_h^2(1-P_{t\bar{t}})=0.27$ that $t$ and $\bar{t}$ cannot be resolved)
shown by blue dashed curve give little sensitivity to $\xi_{htt}$ when
$|\xi_{htt}|\leq\pi/4$. With the inclusion of $t$ or $\bar{t}$
semi-leptonic decay angular correlations, the sensitivity shown by the
red dash-dotted curve shoots above $1\sigma$ at
$|\xi_{htt}|=\pi/4$. Moreover with inclusion of the hadronic decay
angular correlations, as shown by green solid curve $\chi^2_{\rm}$ grows
above 5$\sigma$.
The topponium contribution is not very significant as may be expected from the relatively small topponium formation cross section at $\sqrt{s}=550$~GeV. It may be worth noting that $\xi_{htt}=0$ and $\pi$ can be resolved at 2$\sigma$ level by using decay angular corrections even without beam polarization. 

In the right panel, we show the results for $(\kappa_{htt}^{\rm ex}, \xi_{htt}^{\rm ex})=(1.38,\pi/4)$. As in the $\sqrt{s}=500$~GeV case, $ht\bar{t}$ distribution has almost no power in resolving CP violation (blue dashed curve). $\chi_{\rm min}^2$ value at $\xi_{htt}=-\xi_{htt}^{\rm ex}=-\pi/4$ becomes 4 (2$\sigma$) if we study lepton angular correlations shown by red dash-dotted curve. It jumps to above 4$\sigma$ once we include hadronic decay angular corrections. We can now hope for a meaningful measurement of CP violation in the top Yukawa coupling at $\sqrt{s}=550$~GeV with dedicated efforts.

\begin{figure}[t]
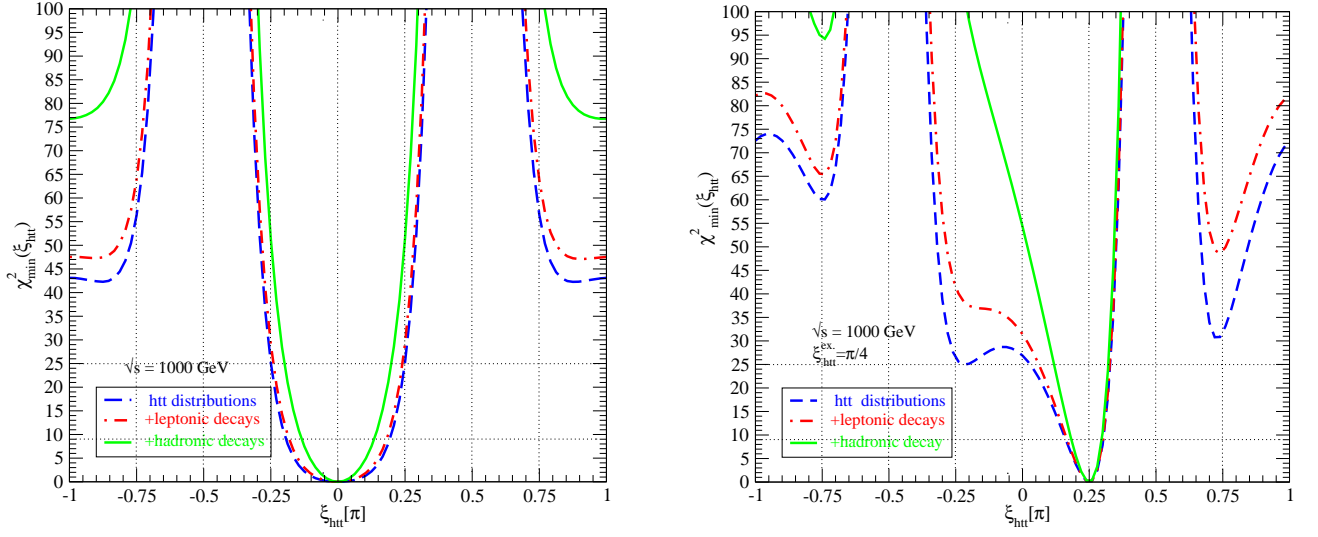

\vspace{0.8cm}
\begin{centering}
\begin{tabular}{c}
\includegraphics[width=0.45\textwidth]{chisq1000_xi0.eps}
\hspace{0.8cm}
\includegraphics[width=0.45\textwidth]{chisq1000_xi025.eps}
\end{tabular}
\caption {The $\chi^2$ plot at $\sqrt{s}=1000$~GeV. $\chi^2_{\rm min}(\xi_{htt})$ is plotted against the fit parameter $\xi_{htt}$, when $\xi_{htt}^{\rm ex}=0$ (the SM Higgs) is assumed for the data (left panel) and when $\xi_{htt}^{\rm ex}=\pi/4$ is assumed (right panel). }
\label{fig:chisq1000}
\end{centering}
\end{figure}
In Fig.~\ref{fig:chisq550CP}, we show the difference 
\begin{eqnarray}
\chi_{\rm min}^2(\xi_{htt}=-\xi_{htt}^{\rm ex})-\chi^2_{\rm min}(\xi_{htt}=\xi_{htt}^{\rm ex})
\label{eq:chimin}
\end{eqnarray}
 as a function of $\xi_{htt}^{\rm ex}$, which gives a measure of the sensitivity of the $e^+e^-\to ht\bar{t}$ experiments to CP violation. {
We believe that by the time the proposed experiment can be done at the ILC we should have relatively strong constraints on the magnitude of $\xi_{htt}$ from the $ht\bar{t}$ and single top$+h$ production cross sections and their distributions at the LHC experiments. The role of ILC experiments should hence be to test that there is indeed CP violation in the Higgs Yukawa couplings, when nonzero magnitude of $\xi_{htt}$ is favored by those data. This can only be done by observing CP violating asymmetries which determine the sign of $\xi_{htt}$, whose significance is proportional to the difference in Eq.~(\ref{eq:chimin}). }
 
{In the figure,} the blue dashed curve shows that the $ht\bar{t}$ distribution has
 almost no sensitivity to CP violation. Inclusion of leptonic decay
 angular correlation (red dash-dotted line) improves, but the most
 significant improvement is found by including hadronic decay angular
 correlations as shown by green solid curve. We find that the contributions from the modes where one
 of $t$ and $\bar{t}$ decays leptonically, and the other decays
 hadronically, whose branching fractions account for $2B_\ell
 B_h\simeq0.44$, increase the $\chi^2_{\rm min}$ value most
 significantly. In this mode, $t$ and $\bar{t}$ are uniquely identified,
 and the leptonic decay angular distribution is exact. Even though the
 hadronic decay angular correlations suffer from $\bar{d}$~v.s.\ $u$~(or
 $\bar{s}$~v.s.\ $c$) misidentification, significant correlation between the $t$
 and $\bar{t}$ decays remains to resolve CP violation. We may tell
 from Fig.~\ref{fig:chisq550CP} that a {2$\sigma$} hint of CP violation
 can be found if $\xi_{htt}>0.12\pi\sim0.38$, while a 3$\sigma$ evidence
 of CP violation can be found if $\xi_{htt}>0.18\pi\sim0.55$ at
 $\sqrt{s}=550$~GeV with $L=1000$~fb$^{-1}$. 

We note in passing that our $\chi^2$ function Eq.~(\ref{eq:chimin})
adopts the experimental distribution $d\sigma_{\rm ex}/d\Phi$ calculated
analytically for $\xi_{htt}=\xi_{htt}^{\rm ex}$ without fluctuation for
realistic binned data. The significance which can be read off from
Fig.~\ref{fig:chisq550CP} should hence be regarded only as a first
optimistic estimate. Dedicated studies with event generation may be
needed to obtain realistic estimates with statistical fluctuation of
experimental data.\footnote{Significance of $\Delta\chi^2$ value
obtained by analytically calculated experimental distribution without
statistical fluctuation is discussed in e.g.\ in section 5 of Ref.~\cite{Ge:2012wj}.}

We close our discussions by repeating the study at $\sqrt{s}=1000$~GeV
with $L=1000$~fb$^{-1}$. The left panel of Fig.~\ref{fig:chisq1000}
shows the constraints on $\xi_{htt}$, when the SM is assumed for the
data. At $\sqrt{s}=1000$~GeV, the $ht\bar{t}$ distribution (blue dashed
curve) has significant sensitivity, and it alone can reject
$|\xi_{htt}|=\pi/4$ at 5$\sigma$ level. The inclusion of both leptonic
and hadronic decay angular correlations (green solid curve) doubles the
$\chi_{\rm min}^2$ value. In the right panel, we show the constraint
when $(\kappa_{htt}^{\rm ex},\xi_{htt}^{\rm ex})=(1.2,\pi/4)$ is assumed for
data. $\chi_{\rm min}^2$ value at $\xi_{htt}=-\xi_{htt}^{\rm ex}=-\pi/4$
tells that CP violation can be seen at 5$\sigma$ even with $ht\bar{t}$ distribution only (blue dashed curve), while it exceeds 10$\sigma$ once we study full decay angular correlations.

\begin{figure}[b]
\vspace{0.8cm}
\begin{centering}
\begin{tabular}{c}
\includegraphics[width=0.45\textwidth]{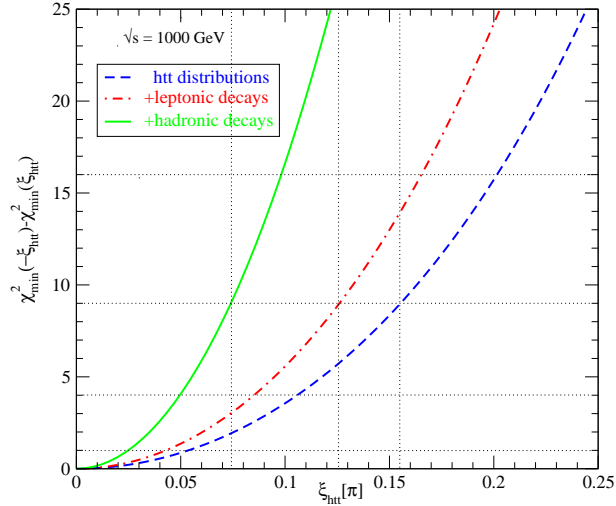}
\end{tabular}
\caption {Sensitivity to CP violation is shown by $\chi_{\rm min}^2 (\xi_{htt}=-\xi_{htt}^{\rm ex})-\chi^2_{\rm min}(\xi_{htt}=\xi_{htt}^{\rm ex})$ plotted against $\xi_{htt}^{\rm ex}$, the assumed true value of the CP phase at $\sqrt{s}=1000$~GeV.}
\label{fig:chisq1000CP}
\end{centering}
\end{figure}

The discovery potential of CP violation at $\sqrt{s} = 1000$~GeV with 1000~fb$^{-1}$ is summarized in Fig.~\ref{fig:chisq1000CP}. From the green solid curve that is obtained from our full correlation studies, we may hope to find a 3$\sigma$ evidence for CP violation if $\xi_{htt}^{\rm ex}\geq0.074\pi\sim0.23$, and a $5\sigma$ discovery if $\xi_{htt}^{\rm ex}\geq0.12\pi\sim0.38$.

\section{summary and discussions}\label{sec:sum}
In this report, we study the potential of future $e^+e^-$ collider experiments to discover CP violation in the top Yukawa coupling, whose magnitude is largest among all the couplings in the SM. We examine $e^+e^-\to ht\bar{t}$ production with all $t$ and $\bar{t}$ decay modes, including both semi-leptonic $(t\to b\bar{\ell}\nu,~\bar{t}\to\bar{b}\ell\bar{\nu})$ and hadronic decays. The full kinematical distributions including the decay angular correlations are obtained by using the density matrix formalism. 
In addition to the well known $t$ and $\bar{t}$ decay density matrices
for semi-leptonic decays, where the full information on the $t$ and
$\bar{t}$ polarization is given by the $\bar{\ell}$ and $\ell$ decay
angular distributions in the $t$ and $\bar{t}$ rest frame respectively,
we introduce novel decay density matrices for the hadronic decay modes
which keep significant $t$ and $\bar{t}$ polarization resolving power
even when one cannot distinguish between $\bar{d}$ v.s.\ $u$ or
$\bar{s}$ v.s.\ $c$ jets ($d$ v.s.\ $\bar{u}$ or $s$ v.s.\ $\bar{c}$ jets) in $W^+~(W^-)$ decays. 

We approximate the full differential cross section by using the decay angular correlations of the leading-order matrix elements with CP-violating $ht\bar{t}$ coupling phase $\xi_{htt}$, corrected for the $m_{t\bar{t}}$ dependent NLO correction factor and the topponium formation at $m_{t\bar{t}}\sim2m_t$. 
{QCD corrections to the angular distributions or correlations 
of top-quark decay products have not been studied in this paper.
Detailed analysis will be required in future work.}
The sensitivity to CP violation has been estimated by comparing the full distributions between $\xi_{htt}^{}$ and $-\xi_{htt}^{}$ in the range $0\leq|\xi_{htt}|<\pi/4$, by adjusting the magnitude of the $ht\bar{t}$ coupling $(\kappa_{htt})$ to reproduce the total cross section of the SM Higgs. 
We study the case at $\sqrt{s}=500$~GeV with ${\cal L}=1000$~fb$^{-1}$, and find that even though the semi-leptonic and hadronic decay angular correlations (as well as topponium contributions) give sensitivity to the sign of $\xi_{htt}$, it reaches only up to the $2\sigma$ level after including all the contributions for the maximum CP phase, $|\xi_{htt}|=\pi/4$.
We therefore examine the possibility of increasing the beam energy by $10\%$ to $\sqrt{s}=550$~GeV. 
The sensitivity grows significantly, giving us the possibility of a
$2\sigma$ hint for $|\xi_{htt}|\sim0.12\pi\sim0.38$ and that of a $3\sigma$ evidence for $|\xi_{htt}|\sim0.18\pi\sim0.55$. 
We therefore propose that the next target energy of a linear $e^+e^-$ collider should be increased from the original 500~GeV to 550~GeV. Impacts of such a $10\%$ increase in the beam energy on other key measurements (such as the $hhh$ coupling) should be studied. 

At $\sqrt{s}=1000$~GeV, we expect a $3\sigma$ evidence at $|\xi_{htt}|\sim0.074\pi\sim0.23$, and a $5\sigma$ discovery at $|\xi_{htt}|\sim0.12\pi\sim0.38$. All these numbers should be regarded as possible maximum sensitivity by a perfect detector that can resolve all information given by the matrix elements.

We believe that the $t$ and $\bar{t}$ decay density matrices we obtain for both semi-leptonic and hadronic decay will be a powerful tool in all processes with $t$ and/or $\bar{t}$ in the final states. Their distributions are explained in detail in Appendix~\ref{sec:appA}. The use of numerical helicity amplitudes evaluated by amplitude calculators can also have wide applications, especially for all the processes whose amplitudes are so complicated that analytic expressions are of little use. We therefore give in Appendix~\ref{sec:appB} a review of the fermion wave function phase convention in Refs.~\cite{Hagiwara:1985yu,Murayama:1992gi} that is adopted in MadGraph~\cite{Stelzer:1994ta}.

\appendix

\section{Top and antitop quark decay density matrix}\label{sec:appA}
In this appendix, we give top quark decay density matrices for 
\begin{eqnarray}
t\to b\bar{\ell}\nu_{\ell}\label{app:lepdecay}
\end{eqnarray}
and
\begin{eqnarray}
t\to b\bar{d}u
\end{eqnarray}
in the top quark rest frame where the $z$-axis is chosen along the top quark momentum $\vec{p}_t$ and the $y$-axis along the $\vec{p}_t\times\vec{p}_h$ direction in the $t\bar{t}$ rest frame (where the $e^+e^-\to ht\bar{t}$ helicity amplitudes are calculated). In the frame, we parameterize the four momentum of $b$ and $\bar{\ell} (\bar{d})$ as 
\begin{eqnarray}
p_b^\mu=\frac{m_t}{2}x_b(1,\sin\theta_b^\ast\cos\phi^\ast_b,\sin\theta_b^\ast\sin\phi_b^\ast,\cos\theta_b^\ast)\label{app:pb}\\
p_{\bar{\ell}(\bar{d})}^\mu=\frac{m_t}{2}\bar{x}(1,\sin\bar{\theta}^\ast\cos\bar{\phi}^\ast,\sin\bar{\theta}^\ast\sin\bar{\phi}^\ast,\cos\bar{\theta}^\ast)\label{app:pl}
\end{eqnarray}
in the limit of neglecting all the final fermion masses including the $b$-quark. 

The decay matrix element for the semileptonic decay (\ref{app:lepdecay}) can be written as 
\begin{eqnarray}\label{eq:A_topamp}
M_\sigma&=&M(t_\sigma\to b\bar{\ell}\nu)\nonumber\\
&=&\frac{g^2}{2}D_W^{}(p_{\bar{\ell}}+p_\nu)~u_L^{}(b)_k^\dagger(\sigma_-^\mu)^{}_{k\ell}u_L^{}(t,\sigma)^\ell~u_L^{}(\nu)_m^\dagger(\sigma_{-\mu})^{}_{mn}v_L^{}(\bar{\ell})_n^{}
\end{eqnarray}
with the $W$ propagator factor, $D_W^{}(q)=1/(q^2-m_W^2+im_W\Gamma_W)$, and $\sigma_{\pm}^\mu=(1,\pm\vec{\sigma})$ are the chiral four vectors of the Pauli matrices. By using the Fierz identity
\begin{eqnarray}
(\sigma_\pm^\mu)_{k\ell}(\sigma_{\pm\mu})_{mn}=2(i\sigma^2)_{km}(i\sigma^2)_{\ell n}
\end{eqnarray}
we can express the amplitudes as
\begin{eqnarray}
M_\sigma=g^2D_W\left\{u_L(b)^\dagger(i\sigma^2)u_L^\ast(\nu)\right\}\left\{u_L(t,\sigma)^T(i\sigma^2)v_L(\bar{\ell})\right\},
\end{eqnarray}
where the bi-fermionic products in each big bracket are separately Lorentz invariant\footnote{The invariance of each bi-spinors is manifest in the spinor rotation
\begin{eqnarray}
&&u_L^{}(b)^\ast_{\dot k}(\sigma_-^\mu)^{\dot k\ell}u_L^{}(t,\sigma)_\ell ~u^L(\nu)_{\dot m}^\ast(\sigma_{-\mu})^{\dot mn}v_L^{}(\bar{\ell})_n^{}\nonumber\\
&=&2u_L^{}(b)^\ast_{\dot k}\epsilon^{\dot k\dot m}u_L^{}(\nu)^\ast_{\dot m}~u_L^{}(t,\sigma)_\ell^{}\epsilon^{\ell n}v_L^{}(\bar{\ell})_n\nonumber,
\end{eqnarray}
where undotted lower indices give left-handed spinors and the dotted indices are for their complex conjugates. $\epsilon^{k\ell}=\epsilon^{\dot k\dot{\ell}}$ are antisymmetric sign factors with $\epsilon_{12}=-\epsilon^{12}=1$. }.

The spinor products are calculated easily as
\begin{eqnarray}
u_L(b)^\dagger(i\sigma^2)u_L^\ast(\nu)=m_{b\nu}=m_t\sqrt{1-\bar{x}}
\end{eqnarray}
in the $b+\nu$ rest frame, while in the $t$ rest frame we find
\begin{eqnarray}
u_L(t,\sigma)^T(i\sigma^2)v_L(\bar{\ell})=m_t\left(\bar{x}\frac{1+\sigma\cos\bar{\theta}^\ast}{2}e^{i\sigma\bar{\phi}^\ast}\right)^{\frac{1}{2}}
\label{app:spin_prod}
\end{eqnarray}
for the top helicity $\sigma/2$. The amplitudes (A7) are hence simply
\begin{eqnarray}
M_\sigma=g^2D_W^{}m_t^2\left(\bar{x}(1-\bar{x})\frac{1+\sigma\cos\bar{\theta}^\ast}{2}{e^{i\sigma\bar{\phi}^\ast}}\right)^{\frac{1}{2}}\label{eq:amp_A}
\end{eqnarray}

We normalize the top quark decay density matrix distribution as 
\begin{eqnarray}
d\rho_{\sigma\sigma^\prime}^{}=\frac{1}{2m_t\Gamma_t}M_\sigma M_{\sigma^\prime}^\ast d\Phi_3(t\to b\bar{\ell}\nu)
\end{eqnarray}
so that the total integral gives
\begin{eqnarray}
\int d\rho_{\sigma\sigma^\prime}=\frac{1}{2m_t\Gamma_t}\int M_\sigma M_{\sigma^\prime}^\ast d\Phi_3(t\to b\bar{\ell}\nu)=\delta_{\sigma\sigma^\prime}B(t\to b\bar{\ell}\nu)
\end{eqnarray}
Note that the trace of (A12) gives twice the branching fraction. Inserting the amplitudes (A10) into (A11), we find
\begin{eqnarray}
d\rho_{\sigma\sigma^\prime}=\frac{g^4m_t^4}{2m_t\Gamma_t}|D_W^{}|^2\bar{x}(1-\bar{x})\left(\frac{1+\sigma\cos\bar{\theta}^\ast}{2}\frac{1+\sigma^\prime\cos\bar{\theta}^\ast}{2}\right)^{\frac{1}{2}}e^{i\frac{\sigma-\sigma^\prime}{2}\bar{\phi}^\ast}d\Phi_3.
\label{eq:A_drhossp}
\end{eqnarray}
In the case of semi-leptonic decays (A1), it is most convenient to parameterize the invariant 3-body phase space as 
\begin{eqnarray}
d\Phi_3=\frac{m_t^2}{128\pi^3}~\bar{x}d\bar{x}~\frac{d\cos\bar{\theta}^\ast d\bar{\phi}^\ast}{4\pi}\frac{d\cos\theta_b^{\ast\ast}}{2}
\end{eqnarray}
where $\theta_b^{\ast\ast}$ denotes the polar angle of the $b$ quark in the $b+\nu$ rest frame measured from the $-\vec{p}_{\bar{\ell}}$ direction. In the zero width limit of the $W$, we can integrate out $\cos\theta_b^{\ast\ast}$ as 
\begin{eqnarray}
&&\int_{-1}^{1}\frac{d\cos\theta_b^{\ast\ast}}{2}~|D_W^{}|^2\nonumber\\
&=&\int_{-1}^1\frac{d\cos\theta_b^{\ast\ast}}{2}\frac{\pi}{m_W^{}\Gamma_W^{}}~\delta(m_t^2\bar{x}\frac{1-\cos\theta_b^{\ast\ast}}{2}-m_W^2)\nonumber\\
&=&\frac{\pi}{m_W^{}\Gamma_W^{}}\frac{1}{m_t^2\bar{x}}~\theta(\bar{x}-\frac{m_W^2}{m_t^2}).
\end{eqnarray}
We integrate out $\bar x$ in the region $m_W^2/m_t^2<\bar{x}<1$, and find 
\begin{eqnarray}
d\rho_{\sigma\sigma^\prime}=B(t\to b\bar{\ell}\nu)~(1+\sigma\cos\bar{\theta}^\ast)^{\frac{1}{2}}~(1+\sigma^\prime\cos\bar{\theta}^\ast)^\frac{1}{2}~e^{i\frac{\sigma-\sigma^\prime}{2}\bar{\phi}^\ast}~\frac{d\cos\bar{\theta}^\ast{d\bar{\phi}^\ast}}{{4}\pi}.
\end{eqnarray}
In the matrix form
\begin{eqnarray}\label{eq:A_drholep}
d\rho=B(t\to b\bar{\ell}\nu)
\left(\begin{array}{cc}
1+\cos\bar{\theta}^\ast&\sin\bar{\theta}^\ast e^{i\bar{\phi}^\ast}\\
\sin\bar{\theta}^\ast e^{-i\bar{\phi}^\ast}&1-\cos\bar{\theta}^\ast
\end{array}\right)
\frac{d\cos\bar{\theta}^\ast d\bar{\phi}^\ast}{4\pi}
\end{eqnarray}
reproducing the normalization Eq.~(A12).

In the case of the hadronic decay (A2) of the top quark, matrix elements are exactly the same as those of the semileptonic decays Eq.~(A10), where $(\bar{x},\bar{\theta}^\ast,\bar{\phi}^\ast)$ are now the normalized energy and the angles of the $\bar{d}$ quark (or $\bar{s}$ in the case of $W\to\bar{s}$c decay); see Eq.~(A4). We need, however, a careful treatment of the decay phase space integral, because it is difficult to identify $\bar{d}$ from $u$ ($\bar{s}$ from $c$) in collider experiments. We introduce a parameter $P_{\bar{d}u}(0\leq P_{\bar{d}u}\leq1)$, which measures the probability that $\bar{d}$ v.s.\ $u$ ($\bar{s}$ v.s.\ $c$) are distinguished correctly, in the decay density matrix distributions. Although all the results presented in this report are for $P_{\bar{d}u}^{}=P_{\bar{s}c}^{}=0$ (absolutely no distinction between the two decaying jets of the $W$), we hope that our formalism encourages further efforts for $\bar{s}$ v.s.\ $c$ discrimination (or even $\bar{d}$ v.s.\ $u$).  

We start with the differential density matrices
Eq.~(\ref{eq:A_drhossp}), multiplied by a factor of 3 for the color,
where ($\bar{x},\bar{\theta}^\ast,\bar{\phi}^\ast$) are parameterizing
$\bar{d}~ (\bar{s})$ quark momentum in the top quark rest frame. Because $\bar{d}$ and $u$ ($\bar{s}$ and $c$) are difficult to distinguish, we parameterize the $t\to b\bar{d}u$ phase space by using the $W\to\bar{d}u$ rest frame angles: 
\begin{eqnarray}
d\Phi_3=\frac{1}{64\pi^2}\left(1-\frac{m_{\bar{d}u}^2}{m_t^2}\right)\frac{d\cos\theta_b^\ast d\phi_b^\ast}{4\pi}\frac{dm_{\bar{d}u}^2}{2\pi}\frac{d\cos\bar{\theta}^{\ast\ast}d\bar{\phi}^{\ast\ast}}{4\pi}
\end{eqnarray}
where $\theta_b^\ast$ and $\phi_b^\ast$ give the $b$ quark momentum in the top quark rest frame, while $\bar{\theta}^{\ast\ast}$ and $\bar{\phi}^{\ast\ast}$ give the $\bar{d}$ momentum in the $W\to\bar{d}u$ rest frame which is obtained from the top quark rest frame by rotations ($-\phi_b^\ast$ about the $z$-axis, and then $-\theta_b^\ast$ about the $y$-axis) and a Lorentz boost along the $z$-axis by 
\begin{eqnarray}
\beta=\frac{1-m_{\bar{d}u}^2/m_t^2}{1+m_{\bar{d}u}^2/m_t^2}.
\end{eqnarray}
The integration over $m_{\bar{d}u}^2$ is done in the zero $W$ width limit to obtain
\begin{eqnarray}
\int^{m_t^2}_0\frac{dm_{\bar{d}u}^2}{2\pi}|D_W^{}|^2=\frac{1}{2m_W^{}\Gamma_W}
\end{eqnarray}
and we find 
\begin{eqnarray}\label{eq:A_rhohad}
d\rho_{\sigma\sigma^\prime}=\frac{6B(t\to b\bar{d}u)}{\left(1-\frac{m_W^2}{m_t^2}\right)\left(1+2\frac{m_W^2}{m_t^2}\right)}\hat{\rho}_{\sigma\sigma^\prime}^{}~
\frac{d\cos\theta_b^\ast d\phi_b^\ast}{4\pi}
\frac{d\cos\bar{\theta}^{\ast\ast}d\bar{\phi}^{\ast\ast}}{4\pi},
\end{eqnarray}
where 
\begin{eqnarray}\label{eq:A_rhob}
\hat{\rho}=(1-\bar{x})
\left(\begin{array}{cc}
\bar{x}+\bar{x}_z&\bar{x}_x+i\bar{x}_y\\
\bar{x}_x-i\bar{x}_y&\bar{x}-\bar{x}_z
\end{array}\right).
\end{eqnarray}
Here, $\bar{x}^\mu=(\bar{x},\bar{x}_x,\bar{x}_y,\bar{x}_z)$ is the
normalized $\bar{d}$ quark four momentum in the top quark rest frame
\begin{eqnarray}\label{eq:A_xbar}
\bar{x}&=&\frac{m_W^{}}{m_t^{}}\gamma^\ast(1-\beta^\ast\cos\bar{\theta}^{\ast\ast})\nonumber\\
\bar{x}_x&=&\frac{m_W^{}}{m_t^{}}\left\{\cos\phi_b^\ast(\cos\theta_b^\ast\sin\bar{\theta}^{\ast\ast}\cos\bar{\phi}^{\ast\ast}+\sin\theta_b^\ast\gamma^\ast(\cos\bar{\theta}^{\ast\ast}-\beta^\ast))-\sin\phi_b^\ast\sin\bar{\theta}^{\ast\ast}\sin\bar{\phi}^{\ast\ast}\right\}\nonumber\\
\bar{x}_y&=&\frac{m_W^{}}{m_t^{}}\left\{\sin\phi_b^\ast(\cos\theta_b^\ast\sin\bar{\theta}^{\ast\ast}\cos\bar{\phi}^{\ast\ast}+\sin\theta_b^\ast\gamma^\ast(\cos\bar{\theta}^{\ast\ast}-\beta^\ast))+\cos\phi_b^\ast\sin\bar{\theta}^{\ast\ast}\sin\bar{\phi}^{\ast\ast}\right\}\nonumber\\
\bar{x}_z&=&\frac{m_W^{}}{m_t^{}}\left\{-\sin\theta_b^\ast\sin\bar{\theta}^{\ast\ast}\cos\bar{\phi}^{\ast\ast}+\cos\theta_b^\ast\gamma^\ast(\cos\bar{\theta}^{\ast\ast}-\beta^\ast)\right\},
\end{eqnarray}
where $\gamma^\ast=\frac{m_t}{2m_W}\left(1+\frac{m_W^2}{m_t^2}\right)$,
$\gamma^\ast\beta^\ast=\frac{m_t}{2m_W}\left(1-\frac{m_W^2}{m_t^2}\right)$. With
the above parameterization, we can express the top quark decay density
matrix distribution under realistic experimental conditions with finite
(or zero) probability $P_{\bar{d}u}$ for $\bar{d}$ v.s.\ $u$ ($\bar{s}$
v.s.\ $c$) discrimination as follows
\begin{eqnarray}\label{eq:drhoP}
d\rho(P_{\bar{d}u})&&=\frac{6B(t\to b\bar{d}u)}{\left(1-\frac{m_W^2}{m_t^2}\right)\left(1+2\frac{m_W^2}{m_t^2}\right)}\nonumber\\
&&\times\left\{\frac{1+P_{\bar{d}u}}{2}\hat{\rho}(\theta_b^\ast,\phi_b^\ast,\bar{\theta}^{\ast\ast},\bar{\phi}^{\ast\ast})+\frac{1-P_{\bar{d}u}}{2}\hat{\rho}(\theta_b^\ast,\phi_b^\ast,\pi-\bar{\theta}^{\ast\ast},\pi+\bar{\phi}^{\ast\ast})\right\}\frac{d\cos\theta_b^\ast d\phi_b^\ast}{4\pi}\frac{d\cos\bar\theta^{\ast\ast}d\bar{\phi}^{\ast\ast}}{4\pi}.
\label{eq:drhoPdu_A}
\end{eqnarray}
 The meaning of Eq.~(\ref{eq:drhoPdu_A}) is that, if $\bar{d}$ v.s.\ $u$ discrimination is perfect, i.e.\ $P_{\bar{d}u}=1$, then the density matrix distribution is exactly the same (up to the branching fraction factors) as that for the semi-leptonic decays (A17).  When $P_{d\bar{u}}=0$, which we assume in our numerical studies, the $\bar{d}$ quark (at $\bar{\theta}^{\ast\ast}$ and $\bar{\phi}^{\ast\ast}$) is not distinguished from the $u$ quark (at $\pi-\bar{\theta}^{\ast\ast}$ and $\pi+\bar{\phi}^{\ast\ast}$) in the $W$ rest frame, and hence their average measures the top quark polarization.
 
 In the case of the $\bar{t}$ decay amplitudes, as compared to Eqs.~(\ref{eq:A_topamp}) and (\ref{app:spin_prod}) , we find
 \begin{eqnarray}
 \overline{M}_{\bar{\sigma}}(\bar{t}_{\bar{\sigma}}\to\bar{b}\ell\bar{\nu})&=&\frac{g^2}{2}D_W^{}(p_\ell+p_{\bar{\nu}})~v_L^\dagger(\bar{t},\bar{\sigma})(\sigma_-^\mu)v_L(\bar{b})~u_L(\ell)^\dagger(\sigma_{-\mu})v_L(\bar{\nu})\\
 &=&g^2D_W\left\{v_L^\dagger(\bar{t},\bar{\sigma})(i\sigma^2)u_L(\ell)^\ast\right\}\left\{v_L(\bar{b})^T(i\sigma^2)v_L(\bar{\nu}))\right\}.
 \end{eqnarray}
 We parameterize the $\bar{b}$ and $\ell (d)$ four momenta exactly in the
 same ways as Eqs.~(\ref{app:pb}, \ref{app:pl}) for $t$ decays
 \begin{eqnarray}
p_{\bar{b}}^\mu=\frac{m_t}{2}x_{\bar{b}}(1,\sin\theta_{\bar{b}}^\ast\cos\phi^\ast_{\bar{b}},\sin\theta_{\bar{b}}^\ast\sin\phi_{\bar b}^\ast,\cos\theta_{\bar{b}}^\ast)\label{app:pb_new}\\
p_{\ell (d)}^\mu=\frac{m_t}{2}x(1,\sin{\theta}^\ast\cos{\phi}^\ast,\sin{\theta}^\ast\sin{\phi}^\ast,\cos{\theta}^\ast)\label{app:pl_new}
\end{eqnarray}
in the $\bar{t}$ rest frame which is obtained from the $t$ rest frame by a Lorentz boost along the $z$-axis. In this way, the azimuthal angles are measured from the common $x$-axis, while the $\bar{t}$ helicity corresponds to its spin component along the negative $z$-axis. In this frame, we find the helicity amplitudes are given as
\begin{eqnarray}
\overline{M}_{\bar{\sigma}}=g^2D_W^{}m_t^2\left(x(1-x)\frac{1+\bar{\sigma}\cos{\theta}^\ast}{2}e^{-i\bar{\sigma}{\phi}^\ast}\right)^{\frac{1}{2}}.\label{eq:ampbar_A}
\end{eqnarray}
Therefore the $\bar{t}$ decay density matrix for the semi-leptonic decays is
  \begin{eqnarray}
 d\bar{\rho}_{}=B(\bar{t}\to\bar{b}\ell\bar{\nu})
  \left(\begin{array}{cc}
{1+\cos\theta^\ast}&{\sin\theta^\ast e^{-i\phi^\ast}}\\
{\sin\theta^\ast e^{i\phi^\ast}}&{1-\cos\theta^\ast}
\end{array}\right)
\frac{d\cos\theta^\ast d\phi^\ast}{4\pi}
\label{eq:drhobar_A}
 \end{eqnarray}
 and that for the hadronic decays reads
  \begin{eqnarray}\label{eq:A_rhobar}
 d\bar{\rho}_{\bar{\sigma}\bar{\sigma}^\prime}=
 \frac{6B(\bar{t}\to\bar{b}d\bar{u})}{(1-\frac{m_W^2}{m_t^2})(1+2\frac{m_W^2}{m_t^2})}
 ~\hat{\bar{\rho}}_{\bar{\sigma}\bar{\sigma}^\prime}~
\frac{d\cos\theta_{\bar{b}}^\ast d\phi_{\bar{b}}^\ast}{4\pi}
\frac{d\cos\theta^{\ast\ast} d\phi^{\ast\ast}}{4\pi}
 \end{eqnarray}
 
 \begin{eqnarray}\label{eq:A_x}
  \hat{\bar{\rho}}=(1-x)
  \left(\begin{array}{cc}
{x+x_z}&{x_x-ix_y}\\
{x_x+ix_y}&{x-x_z}
\end{array}\right),
 \end{eqnarray}
 if we can distinguish between $d$ and $\bar{u}$ jets. As in
 Eq.~(\ref{eq:A_xbar}), $x^\mu=(x,x_x,x_y,x_z)$ is the normalized $d$
 (or $s$) quark four momentum in the $\bar{t}$ rest frame 
 \begin{eqnarray}
 x&=&\frac{m_W^{}}{m_t^{}}\gamma^\ast(1-\beta^\ast\cos{\theta}^{\ast\ast})\nonumber\\
 x_x&=&\frac{m_W^{}}{m_t^{}}\left\{\cos\phi_{\bar{b}}^\ast(\cos\theta_{\bar{b}}^\ast\sin{\theta}^{\ast\ast}\cos{\phi}^{\ast\ast}+\sin\theta_{\bar{b}}^\ast\gamma^\ast(\cos{\theta}^{\ast\ast}-\beta^\ast))-\sin\phi_{\bar{b}}^\ast\sin{\theta}^{\ast\ast}\sin{\phi}^{\ast\ast}\right\}\nonumber\\
 x_y&=&\frac{m_W^{}}{m_t^{}}\left\{\sin\phi_{\bar{b}}^\ast(\cos\theta_{\bar{b}}^\ast\sin{\theta}^{\ast\ast}\cos{\phi}^{\ast\ast}+\sin\theta_{\bar{b}}^\ast\gamma^\ast(\cos{\theta}^{\ast\ast}-\beta^\ast))+\cos\phi_{\bar{b}}^\ast\sin{\theta}^{\ast\ast}\sin{\phi}^{\ast\ast}\right\}\nonumber\\
 x_z&=&\frac{m_W^{}}{m_t^{}}\left\{-\sin\theta_{\bar{b}}^\ast\sin{\theta}^{\ast\ast}\cos{\phi}^{\ast\ast}+\cos\theta_{\bar{b}}^\ast\gamma^\ast(\cos{\theta}^{\ast\ast}-\beta^\ast)\right\}.
 \end{eqnarray}
 With the probability $P_{d\bar{u}}=P_{\bar{d}u}$ for $d$ v.s.\ $\bar{u}$ discrimination, the $\bar{t}$ decay density matrix for hadronic decays becomes 
 \begin{eqnarray}
d\bar{\rho}(P_{d\bar{u}})&&=\frac{6B(\bar{t}\to \bar{b}\bar{d}u)}{\left(1-\frac{m_W^2}{m_t^2}\right)\left(1+2\frac{m_W^2}{m_t^2}\right)}\nonumber\\
&&\times\left\{\frac{1+P_{d\bar{u}}}{2}\hat{\bar{\rho}}(\theta_{\bar{b}}^\ast,\phi_{\bar{b}}^\ast,\theta^{\ast\ast},{\phi}^{\ast\ast})+\frac{1-P_{d\bar{u}}}{2}\hat{\bar{\rho}}(\theta_{\bar{b}}^\ast,\phi_{\bar{b}}^\ast,\pi-{\theta}^{\ast\ast},\pi+{\phi}^{\ast\ast})\right\}\frac{d\cos\theta_{\bar{b}}^\ast d\phi_{\bar{b}}^\ast}{4\pi}\frac{d\cos\theta^{\ast\ast}d{\phi}^{\ast\ast}}{4\pi}.
 \end{eqnarray}

It is worth noting here that the $t\to b\bar{\ell}\nu$ and
$\bar{t}\to\bar{b}\ell\bar{\nu}$ amplitudes as shown in
Eq.~(\ref{eq:amp_A}) and Eq.~(\ref{eq:ampbar_A}), respectively, are
expressed as  squared roots of complex numbers because space rotations
on spinors give half angles. These elegant expressions cannot be 
obtained if one adopts a phase convention that depends on the azimuthal
angle, such as the one in Refs.~\cite{Hagiwara:1985yu, Murayama:1992gi}
adopted by a Feynman amplitude calculator
MadGraph~\cite{Stelzer:1994ta}. The amplitudes Eq.~(\ref{eq:amp_A}) and
Eq.~(\ref{eq:ampbar_A}) can be recovered by supplying the phase factor
in Eq.~(\ref{eq:uv_B}), as will be explained in detail in
Appendix~\ref{sec:appB}. Because the phase factor associated with the
$\bar{\ell}$ and $\ell$ spinors, respectively, for $t$ and $\bar{t}$
decay amplitudes are common for both helicity amplitudes, the decay
density matrices in Eqs.~(\ref{eq:A_drholep}, \ref{eq:A_rhob},
\ref{eq:drhobar_A}, \ref{eq:A_x}) are independent of the $\bar{\ell}$  and $\ell$ spinor phase convention.

\section{ HELAS phase convention for the density matrix}\label{sec:appB}

As explained above, we use HELAS amplitudes as obtained by MadGraph to test our analytic calculations, and to generate differential cross sections with full decay correlations. We encounter a subtle frame dependence of the $ht\bar{t}$ production density matrix elements due to the specific phase convention, hereafter called HZ convention~\cite{Hagiwara:1985yu}, which has been adopted by HELAS subroutines~\cite{Murayama:1992gi}. Because the use of MadGraph-generated HELAS amplitudes can be a powerful tool for studying massive fermion spin correlations in various experiments, we show in this appendix the origin of the frame dependence and ways to avoid possible inconsistencies. 

In Ref.~\cite{Hagiwara:1985yu}, massive fermion wave functions are expressed in terms of two helicity spinors 
\begin{subequations}
\begin{align}
\chi_+(\vec{p})&=\left(\begin{array}{c}
\cos\frac{\theta}{2}\\
\sin\frac{\theta}{2}e^{i\phi}
\end{array}\right)\\
\chi_-(\vec{p})&=\left(\begin{array}{c}
-\sin\frac{\theta}{2}e^{-i\phi}\\
\cos\frac{\theta}{2}
\end{array}\right)
\end{align}
\end{subequations}
when the fermion four momentum is 
\begin{eqnarray}\label{appB:2pmu}
p^\mu=(E,\vec{p})=(E,p\sin\theta\cos\phi,p\sin\theta\sin\phi,p\cos\theta)
\end{eqnarray}
where $p=|\vec{p}|$, $0\leq\theta<\pi$ so that $\cos\frac{\theta}{2},\sin\frac{\theta}{2}>0$ and $0\leq\phi<2\pi$. Note that 
\begin{subequations}
\begin{align}
\chi_-(\vec{p})&=-i\sigma^2\chi_+(\vec{p})^\ast\\
\chi_+(\vec{p})&=i\sigma^2\chi_-(\vec{p})^\ast.
\end{align}
\end{subequations}
By starting from the fermion spinors in the rest frame, 
\begin{eqnarray}
u_L^{}(p,+)=u_R^{}(p,+)=\sqrt{m}
\left(\begin{array}{c}
1\\
0
\end{array}\right),
~~~~~
u_L^{}(p,-)=u_R^{}(p,-)=\sqrt{m}
\left(\begin{array}{c}
0\\
1
\end{array}\right),
\end{eqnarray}
and with the charge conjugation 
\begin{eqnarray}
v_L^{}(p,\sigma)=i\sigma^2 u_R^{}(p,\sigma)^\ast,
~~~~
v_{R}^{}(p,\sigma)=-i\sigma^2u_L(p,\sigma)^\ast,
\end{eqnarray}
we find that straightforward calculation of the Lorentz transformation, from the $t$ and $(\bar{t})$ rest frame $p^\mu=(m,0,0,0)$ to the frame in which the $t(\bar{t})$ four momentum becomes Eq.~(\ref{appB:2pmu}),
\begin{subequations}
\begin{align}
u(p,+)_{\rm}
&=\left(\begin{array}{c}
\sqrt{E-p}~\chi_+(\vec{p})\\
\sqrt{E+p}~\chi_+(\vec{p})
\end{array}\right)e^{-i\frac{\phi}{2}}\\
u(p,-)_{}
&=\left(\begin{array}{c}
\sqrt{E+p}~\chi_-(\vec{p})\\
\sqrt{E-p}~\chi_-(\vec{p})
\end{array}\right)e^{i\frac{\phi}{2}}\\
v(p,+)_{\rm}
&=\left(\begin{array}{c}
-\sqrt{E+p}~\chi_-(\vec{p})\\
\sqrt{E-p}~\chi_-(\vec{p})
\end{array}\right)e^{i\frac{\phi}{2}}\\
v(p,-)_{\rm}
&=\left(\begin{array}{c}
\sqrt{E-p}~\chi_+(\vec{p})\\
-\sqrt{E+p}~\chi_+(\vec{p})
\end{array}\right)e^{-i\frac{\phi}{2}}
\end{align}
\end{subequations}
The fermion wave functions in Ref.~\cite{Hagiwara:1985yu}, adopted in HELAS~\cite{Murayama:1992gi} are obtained by chopping off the phase factors which depend on the azimuthal angle and the helicity of $t$ and $\bar{t}$,
\begin{subequations}
\begin{align}
u(p,\sigma)&=u(p,\sigma)_{\rm HZ}^{}~e^{-i\frac{\sigma\phi}{2}}\\
v(p,\bar{\sigma})&=v(p,\bar{\sigma})_{\rm HZ}^{}~{e^{i\frac{\bar{\sigma}\phi}{2}}}
\end{align}
\label{eq:uv_B}
\end{subequations}
\hspace{-0.18cm}
Hence, for all $t\bar{t}$ production processes, the helicity amplitudes in the HZ phase convention differ from the naive ones as 
\begin{eqnarray}
M_{\sigma\bar{\sigma}}=(M_{\sigma\bar{\sigma}})_{\rm HZ}~(e^{i\frac{\sigma\phi+\bar{\sigma}\bar{\phi}}{2}}),
\label{app:amp_phase}
\end{eqnarray}
when $t$ and $\bar{t}$ momenta have non-zero azimuthal angles, $\phi$ and $\bar{\phi}$, respectively.

In our study, we calculate the helicity amplitudes in the $t\bar{t}$ rest frame where
\begin{eqnarray}
\bar{\phi}=\phi
\end{eqnarray}
holds.\footnote{In order to avoid purely complex phase $e^{\pm i\pi/2}$ for the $\bar{t}$ wave function, we obtain the $\bar{t}$ momentum by setting $\theta_{\bar{t}}=-\pi+\theta_t$ and $\phi_{\bar{t}}=\phi_t$ in the $t\bar{t}$ rest frame.}
 Therefore the helicity amplitudes in HZ convention (the HELAS amplitudes) differ from the conventional amplitudes by the following phase factors
\begin{eqnarray}
M_{++}&=&(M_{++})_{\rm HZ}^{}~e^{i\phi}\nonumber\\
M_{--}&=&(M_{--})_{\rm HZ}^{}~e^{-i\phi}\nonumber\\
M_{\pm\mp}&=&(M_{\pm\mp})_{\rm HZ}^{}.
\end{eqnarray}
Although the phase factors do not affect the magnitudes of each helicity
amplitudes, they do not decouple from the off-diagonal elements of the
density matrix $\rho^{}_{\sigma\bar{\sigma},\sigma^\prime\bar\sigma^\prime}=M_{\sigma\bar\sigma}^{}M_{\sigma'\bar\sigma'}^\ast$, 
\begin{eqnarray}
\rho_{++,--}&=&(\rho_{++,--})_{\rm HZ}^{}~e^{i2\phi}\nonumber\\
\rho_{++,\pm\mp}&=&(\rho_{++,\pm\mp})_{\rm HZ}^{}~e^{i\phi}\nonumber\\
\rho_{+-,-+}&=&(\rho_{+-,-+})_{\rm HZ}^{}.~
\label{app:rhoHZ}
\end{eqnarray}
They introduce unphysical $\phi$-dependences in real and imaginary parts of the density matrices, when $t$ and $\bar{t}$ momentum has non-zero azimuthal angles. {Note that the phase does not appear in the chirality conserving amplitudes in the massless limit, and hence the $e^+e^-$ currents are free from the azimuthal angle phases. }

Such convention dependent $\phi$-dependence does not survive in the physical distributions. For instance, the phases in 
Eq.~(\ref{app:rhoHZ}) are precisely canceled by the corresponding phases in the $t$ and $\bar{t}$ decay density matrices, if we use the same wave functions for production and decay. In order to recover the rotational invariance of the conventional helicity amplitudes, we may supply the phase factor in Eq.~(\ref{app:amp_phase}) or simply stick to the frame where $t$ and $\bar{t}$ momenta have no $y$-components. We adopt in Section~\ref{sec:helicity} the latter approach by choosing the frame in which $t$ and $\bar{t}$ momenta are in the $x$-$z$ plane, whereas the azimuthal angles are given to $e$ and $\bar{e}$ momenta. 

\section*{Acknowledgement}

The authors wish to thank Pyungwon Ko in KIAS, Tao Han in University of Pittsburgh and Shinya Kanemura in University of Toyama and Osaka University for support where part of this work was done. YZ thanks Kentarou Mawatari for helpful discussions and JSPS for financial support. This work is supported in part by Grant-in-Aid for Scientific Research (No. 16F16321) from JSPS.


\begin{thebibliography}{99}

 %\cite{Aad:2015gba}
\bibitem{Aad:2015gba} 
  G.~Aad {\it et al.} [ATLAS Collaboration],
  %``Measurements of the Higgs boson production and decay rates and coupling strengths using pp collision data at $\sqrt{s}=7$ and 8 TeV in the ATLAS experiment,''
  Eur.\ Phys.\ J.\ C {\bf 76}, no. 1, 6 (2016).
%  doi:10.1140/epjc/s10052-015-3769-y
%  [arXiv:1507.04548 [hep-ex]].
%%CITATION = doi:10.1140/epjc/s10052-015-3769-y;%%

%\cite{Khachatryan:2016vau}
\bibitem{Khachatryan:2016vau} 
  G.~Aad {\it et al.} [ATLAS and CMS Collaborations],
  %``Measurements of the Higgs boson production and decay rates and constraints on its couplings from a combined ATLAS and CMS analysis of the LHC pp collision data at $ \sqrt{s}=7 $ and 8 TeV,''
  JHEP {\bf 1608}, 045 (2016).
%  doi:10.1007/JHEP08(2016)045
%  [arXiv:1606.02266 [hep-ex]].
%%CITATION = doi:10.1007/JHEP08(2016)045;%%

{
%\cite{Gunion:1996xu}
\bibitem{Gunion:1996xu}
  J.~F.~Gunion and X.~G.~He,
  %``Determining the CP nature of a neutral Higgs boson at the LHC,''
  Phys.\ Rev.\ Lett.\  {\bf 76} (1996) 4468.
%  doi:10.1103/PhysRevLett.76.4468
  %[hep-ph/9602226].
  %%CITATION = doi:10.1103/PhysRevLett.76.4468;%%
  %56 citations counted in INSPIRE as of 09 Feb 2018
}

{
  %\cite{Ellis:2013yxa}
\bibitem{Ellis:2013yxa}
  J.~Ellis, D.~S.~Hwang, K.~Sakurai and M.~Takeuchi,
  %``Disentangling Higgs-Top Couplings in Associated Production,''
  JHEP {\bf 1404} (2014) 004.
 % doi:10.1007/JHEP04(2014)004
  %[arXiv:1312.5736 [hep-ph]].
  %%CITATION = doi:10.1007/JHEP04(2014)004;%%
  %76 citations counted in INSPIRE as of 09 Feb 2018
  }
 {
  %\cite{Demartin:2014fia}
\bibitem{Demartin:2014fia}
  F.~Demartin, F.~Maltoni, K.~Mawatari, B.~Page and M.~Zaro,
  %``Higgs characterisation at NLO in QCD: CP properties of the top-quark Yukawa interaction,''
  Eur.\ Phys.\ J.\ C {\bf 74} (2014) no.9,  3065.
 % doi:10.1140/epjc/s10052-014-3065-2
  %[arXiv:1407.5089 [hep-ph]].
  %%CITATION = doi:10.1140/epjc/s10052-014-3065-2;%%
  %67 citations counted in INSPIRE as of 11 Feb 2018
  }
 
 %\cite{Khatibi:2014bsa}
\bibitem{Khatibi:2014bsa}
  S.~Khatibi and M.~Mohammadi Najafabadi,
  %``Exploring the Anomalous Higgs-top Couplings,''
  Phys.\ Rev.\ D {\bf 90} (2014) no.7,  074014.
 % doi:10.1103/PhysRevD.90.074014
  %[arXiv:1409.6553 [hep-ph]].
  %%CITATION = doi:10.1103/PhysRevD.90.074014;%%
  %34 citations counted in INSPIRE as of 19 Feb 2018
   
{
%\cite{He:2014xla}
\bibitem{He:2014xla}
  X.~G.~He, G.~N.~Li and Y.~J.~Zheng,
  %``Probing Higgs boson $CP$ Properties with $t\bar{t}H$ at the LHC and the 100 TeV $pp$ collider,''
  Int.\ J.\ Mod.\ Phys.\ A {\bf 30} (2015) no.25,  1550156.
%  doi:10.1142/S0217751X15501560
  %[arXiv:1501.00012 [hep-ph]].
  %%CITATION = doi:10.1142/S0217751X15501560;%%
  %23 citations counted in INSPIRE as of 09 Feb 2018
}

{
%\cite{Buckley:2015vsa}
\bibitem{Buckley:2015vsa}
  M.~R.~Buckley and D.~Goncalves,
  %``Boosting the Direct CP Measurement of the Higgs-Top Coupling,''
  Phys.\ Rev.\ Lett.\  {\bf 116} (2016) no.9,  091801.
  %doi:10.1103/PhysRevLett.116.091801
  %[arXiv:1507.07926 [hep-ph]].
  %%CITATION = doi:10.1103/PhysRevLett.116.091801;%%
  %41 citations counted in INSPIRE as of 09 Feb 2018
  }


{
%\cite{Biswas:2012bd}
\bibitem{Biswas:2012bd}
  S.~Biswas, E.~Gabrielli and B.~Mele,
  %``Single top and Higgs associated production as a probe of the Htt coupling sign at the LHC,''
  JHEP {\bf 1301} (2013) 088.
 % doi:10.1007/JHEP01(2013)088
  %[arXiv:1211.0499 [hep-ph]].
  %%CITATION = doi:10.1007/JHEP01(2013)088;%%
  %70 citations counted in INSPIRE as of 09 Feb 2018
}


  
{  
%\cite{Englert:2014pja}
\bibitem{Englert:2014pja}
  C.~Englert and E.~Re,
  %``Bounding the top Yukawa coupling with Higgs-associated single-top production,''
  Phys.\ Rev.\ D {\bf 89} (2014) no.7,  073020.
 % doi:10.1103/PhysRevD.89.073020
  %[arXiv:1402.0445 [hep-ph]].
  %%CITATION = doi:10.1103/PhysRevD.89.073020;%%
  %28 citations counted in INSPIRE as of 09 Feb 2018
}


%\cite{Demartin:2015uha}
\bibitem{Demartin:2015uha} 
  F.~Demartin, F.~Maltoni, K.~Mawatari and M.~Zaro,
  %``Higgs production in association with a single top quark at the LHC,''
  Eur.\ Phys.\ J.\ C {\bf 75}, no. 6, 267 (2015).
%  doi:10.1140/epjc/s10052-015-3475-9
%  [arXiv:1504.00611 [hep-ph]].
%%CITATION = doi:10.1140/epjc/s10052-015-3475-9;%%

{
 %\cite{Bhattacharyya:2012tj}
\bibitem{Bhattacharyya:2012tj}
  G.~Bhattacharyya, D.~Das and P.~B.~Pal,
  %``Modified Higgs couplings and unitarity violation,''
  Phys.\ Rev.\ D {\bf 87} (2013) 011702.
 % doi:10.1103/PhysRevD.87.011702
  %[arXiv:1212.4651 [hep-ph]].
  %%CITATION = doi:10.1103/PhysRevD.87.011702;%%
  %27 citations counted in INSPIRE as of 09 Feb 2018
 }

{ 
  %\cite{Choudhury:2012tk}
\bibitem{Choudhury:2012tk}
  D.~Choudhury, R.~Islam and A.~Kundu,
  %``Anomalous Higgs Couplings as a Window to New Physics,''
  Phys.\ Rev.\ D {\bf 88} (2013) no.1,  013014.
 % doi:10.1103/PhysRevD.88.013014
  %[arXiv:1212.4652 [hep-ph]].
  %%CITATION = doi:10.1103/PhysRevD.88.013014;%%
  %24 citations counted in INSPIRE as of 09 Feb 2018
}

{
%\cite{BarShalom:1995jb}
\bibitem{BarShalom:1995jb}
  S.~Bar-Shalom, D.~Atwood, G.~Eilam, R.~R.~Mendel and A.~Soni,
  %``Large tree level CP violation in $e^{+} e^{-} \to t \bar{t} H^0$ in the two Higgs doublet model,''
  Phys.\ Rev.\ D {\bf 53} (1996) 1162.
%  doi:10.1103/PhysRevD.53.1162
 % [hep-ph/9508314].
  %%CITATION = doi:10.1103/PhysRevD.53.1162;%%
  %43 citations counted in INSPIRE as of 09 Feb 2018
}
{
%\cite{BhupalDev:2007ftb}
\bibitem{BhupalDev:2007ftb}
  P.~S.~Bhupal Dev, A.~Djouadi, R.~M.~Godbole, M.~M.~Muhlleitner and S.~D.~Rindani,
  %``Determining the CP properties of the Higgs boson,''
  Phys.\ Rev.\ Lett.\  {\bf 100} (2008) 051801.
  %doi:10.1103/PhysRevLett.100.051801
  %[arXiv:0707.2878 [hep-ph]].
  %%CITATION = doi:10.1103/PhysRevLett.100.051801;%%
  %80 citations counted in INSPIRE as of 09 Feb 2018
}
  
   %\cite{Hagiwara:2016rdv}
\bibitem{Hagiwara:2016rdv}
  K.~Hagiwara, K.~Ma and H.~Yokoya,
  %``Probing CP violation in $e^{+}e^{-}$ production of the Higgs boson and toponia,''
  JHEP {\bf 1606}, 048 (2016).
%  doi:10.1007/JHEP06(2016)048
%  [arXiv:1602.00684 [hep-ph]].
  %%CITATION = doi:10.1007/JHEP06(2016)048;%%
  %2 citations counted in INSPIRE as of 17 Feb 2017
  
%\cite{Wolfenstein:1964ks}
\bibitem{Wolfenstein:1964ks}
  L.~Wolfenstein,
  %``Violation of CP Invariance and the Possibility of Very Weak Interactions,''
  Phys.\ Rev.\ Lett.\  {\bf 13}, 562 (1964).
%  doi:10.1103/PhysRevLett.13.562
  %%CITATION = doi:10.1103/PhysRevLett.13.562;%%
  %830 citations counted in INSPIRE as of 27 Nov 2017

%\cite{Alwall:2014hca}
\bibitem{Alwall:2014hca} 
  J.~Alwall {\it et al.},
  %``The automated computation of tree-level and next-to-leading order differential cross sections, and their matching to parton shower simulations,''
  JHEP {\bf 1407}, 079 (2014).
%  doi:10.1007/JHEP07(2014)079
%  [arXiv:1405.0301 [hep-ph]].
  %%CITATION = doi:10.1007/JHEP07(2014)079;%%
  %1955 citations counted in INSPIRE as of 21 Jun 2017

%\cite{Farrell:2005fk}
\bibitem{Farrell:2005fk} 
  C.~Farrell and A.~H.~Hoang,
  %``The Large Higgs energy region in Higgs associated top pair production at the linear collider,''
  Phys.\ Rev.\ D {\bf 72}, 014007 (2005).
%  doi:10.1103/PhysRevD.72.014007
%  [hep-ph/0504220].
%%CITATION = doi:10.1103/PhysRevD.72.014007;%%

%\cite{Farrell:2006xe}
\bibitem{Farrell:2006xe} 
  C.~Farrell and A.~H.~Hoang,
  %``Next-to-leading-logarithmic QCD corrections to the cross- section sigma(e+ e- ---> t anti-t H) at 500-GeV,''
  Phys.\ Rev.\ D {\bf 74}, 014008 (2006).
%  doi:10.1103/PhysRevD.74.014008
%  [hep-ph/0604166].
  %%CITATION = doi:10.1103/PhysRevD.74.014008;%%
  %20 citations counted in INSPIRE as of 21 Jun 2017

%\cite{Yonamine:2011jg}
\bibitem{Yonamine:2011jg} 
  R.~Yonamine, K.~Ikematsu, T.~Tanabe, K.~Fujii, Y.~Kiyo, Y.~Sumino and H.~Yokoya,
  %``Measuring the top Yukawa coupling at the ILC at $\sqrt{s}=500$ GeV,''
  Phys.\ Rev.\ D {\bf 84}, 014033 (2011).
%  doi:10.1103/PhysRevD.84.014033
%  [arXiv:1104.5132 [hep-ph]].
%%CITATION = doi:10.1103/PhysRevD.84.014033;%%

%\cite{Sumino:2010bv}
\bibitem{Sumino:2010bv} 
  Y.~Sumino and H.~Yokoya,
  %``Bound-state effects on kinematical distributions of top quarks at hadron colliders,''
  JHEP {\bf 1009}, 034 (2010)
  Erratum: [JHEP {\bf 1606}, 037 (2016)].
%  doi:10.1007/JHEP06(2016)037, 10.1007/JHEP09(2010)034
%  [arXiv:1007.0075 [hep-ph]].
%%CITATION = doi:10.1007/JHEP06(2016)037, 10.1007/JHEP09(2010)034;%%  
 

%\cite{Ge:2012wj}
\bibitem{Ge:2012wj}
  S.~F.~Ge, K.~Hagiwara, N.~Okamura and Y.~Takaesu,
  %``Determination of mass hierarchy with medium baseline reactor neutrino experiments,''
  JHEP {\bf 1305}, 131 (2013).
%  doi:10.1007/JHEP05(2013)131
%  [arXiv:1210.8141 [hep-ph]].
  %%CITATION = doi:10.1007/JHEP05(2013)131;%%
  %58 citations counted in INSPIRE as of 15 Dec 2017
  
      %\cite{Hagiwara:1985yu}
\bibitem{Hagiwara:1985yu} 
  K.~Hagiwara and D.~Zeppenfeld,
  %``Helicity Amplitudes for Heavy Lepton Production in e+ e- Annihilation,''
  Nucl.\ Phys.\ B {\bf 274}, 1 (1986).
%  doi:10.1016/0550-3213(86)90615-2
  %%CITATION = doi:10.1016/0550-3213(86)90615-2;%%
  %303 citations counted in INSPIRE as of 28 Nov 2017
  
  \bibitem{Murayama:1992gi} 
  %\cite{Hagiwara:1990dw}
%\bibitem{Hagiwara:1990dw}
  K.~Hagiwara, H.~Murayama and I.~Watanabe,
  %``Search for the Yukawa interaction in the process e+ e- ---> t anti-t Z at TeV linear colliders,''
  Nucl.\ Phys.\ B {\bf 367}, 257 (1991).
%  doi:10.1016/0550-3213(91)90017-R
  %%CITATION = doi:10.1016/0550-3213(91)90017-R;%%
  %76 citations counted in INSPIRE as of 15 Dec 2017
  %\cite{Murayama:1992gi}
  H.~Murayama, I.~Watanabe and K.~Hagiwara,
  %``HELAS: HELicity amplitude subroutines for Feynman diagram evaluations,''
  KEK-91-11.
  %%CITATION = KEK-91-11;%%
  %59 citations counted in INSPIRE as of 28 Nov 2017
  
%\cite{Stelzer:1994ta}
\bibitem{Stelzer:1994ta}
  T.~Stelzer and W.~F.~Long,
  %``Automatic generation of tree level helicity amplitudes,''
  Comput.\ Phys.\ Commun.\  {\bf 81}, 357 (1994).
%  doi:10.1016/0010-4655(94)90084-1
%  [hep-ph/9401258].
  %%CITATION = doi:10.1016/0010-4655(94)90084-1;%%
  %907 citations counted in INSPIRE as of 15 Dec 2017


  
\end{thebibliography}
\end{document}